\documentclass[final,onecolumn,11pt]{IEEEtran}

\usepackage[utf8]{inputenc}
\usepackage{graphicx}
\usepackage{amssymb}
\usepackage[cmex10]{amsmath}
\usepackage{color}
\usepackage{theorem}
\usepackage{hyperref}
\usepackage{cite}
\usepackage{url}
\usepackage{breakurl}
\usepackage[english]{babel}
\usepackage[english = american]{csquotes}
\usepackage{mathtools}
\MakeOuterQuote{"}

\definecolor{institut_color_orig}{rgb}{0.8,0.3,0.4}
\definecolor{blue_aw}{rgb}{0.3,0.3,1.0}

\usepackage{pgf}
\usepackage{tikz}
\usetikzlibrary{matrix,chains,positioning,arrows,calc,decorations,plotmarks,patterns, fit,backgrounds}
\usepackage{pgfplots}
\usetikzlibrary{external}                       
\tikzsetexternalprefix{tikzpic/}                   

\pgfplotscreateplotcyclelist{mylist}{red,blue,black,yellow,brown}

\tikzset{
    >=stealth',
    mycircle/.style={circle, draw=gray, very thick, text width=.1em, minimum height=1.5em, text centered},
    mycircle_small/.style={circle,draw=awgray_dark,fill = awgray_dark, inner sep=0,minimum size=.6em},
    mycircle_small_black/.style={circle,draw=black,fill = black, inner sep=0,minimum size=.6em},
    mybox/.style={rectangle,rounded corners,draw=black, thick,text width=1em,minimum height=4em,minimum width=4em,text centered},
    mybox_small/.style={rectangle,rounded corners,draw=black, thick,text width=1em,minimum height=2em,minimum width=2em,text centered},
    mybox_vec/.style={rectangle,rounded corners,draw=black, thick,text width=1em,minimum height=0.7em, minimum width=4em,text centered},
    mybox_vec_short/.style={rectangle,rounded corners,draw=black, thick,text width=1em,minimum height=0.7em, minimum width=2em,text centered},
    pil/.style={->, thick, shorten <=2pt, shorten >=2pt,},
}

\newtheorem{theorem}{Theorem}
\newtheorem{definition}{Definition}
\newtheorem{lemma}{Lemma}

\newtheorem{problem}{Problem}

\newtheorem{example}{Example}

\newtheorem{construction}{Construction}

\newcommand{\Fqm}{\ensuremath{\mathbb F_{q^m}}}

\newcommand{\Fq}{\ensuremath{\mathbb F_{q}}}
\newcommand{\Fqext}[1]{\ensuremath{\mathbb F_{q^{#1}}}}

\newcommand{\intervallincl}[2]{\ensuremath{[#1,#2]}}
\newcommand{\intervallexcl}[2]{\ensuremath{[#1,#2-1]}}

\newcommand{\NormbasisOrdered}{\boldsymbol{\Normelement}}

\newcommand{\Normelement}{\beta}

\newcommand{\extsmallfielddeg}[1]{\ensuremath{\psi}_{#1}}

\newcommand{\extsmallfieldinputdeg}[2]{\ensuremath{\psi_{#2}\left(#1\right)}}

\newcommand{\fontmetric}[1]{\mathsf{#1}}
\DeclareMathOperator{\defi}{def}
\newcommand{\defeq}{\overset{\defi}{=}}

\DeclareMathOperator{\wt}{wt}
\newcommand{\wtH}[1]{\wt_{\fontmetric{H}}(#1)}

\DeclareMathOperator{\rk}{rk}

\renewcommand{\vec}[1]{\ensuremath{\mathbf{#1}}}
\newcommand{\Mat}[1]{\ensuremath{\mathbf{#1}}}
\newcommand{\vecelements}[1]{\ensuremath{(#1_0 \ #1_1 \ \dots \ #1_{n-1})}}
\newcommand{\vecelementsm}[1]{\ensuremath{(#1_0 \ #1_1 \ \dots \ #1_{m-1})}}

\newcommand{\MoormatExplicitIncl}[3]{
\begin{pmatrix}
#1_{0} & #1_{1} & \dots& #1_{#3-1}\\
#1_{0}^{[1]} & #1_{1}^{[1]} & \dots& #1_{#3-1}^{[1]}\\
\vdots &\vdots&\ddots& \vdots\\
#1_{0}^{[#2]} & #1_{1}^{[#2]} & \dots& #1_{#3-1}^{[#2]}\\
\end{pmatrix}}

\renewcommand{\a}{\vec{a}}
\renewcommand{\b}{\vec{b}}
\renewcommand{\c}{\vec{c}}

\newcommand{\A}{\Mat{A}}
\newcommand{\B}{\Mat{B}}
\newcommand{\C}{\Mat{C}}
\newcommand{\D}{\mathbf D}

\newcommand{\G}{\mathbf G}

\newcommand{\0}{\vec{0}}

\newcommand{\mycode}[1]{\ensuremath{\mathcal{#1}}}
\newcommand{\mycodeRank}[1]{\ensuremath{\mathbb{#1}}}

\newcommand{\codelinear}[1]{\ensuremath{[#1]_q^\fontmetric{R}}}

\newcommand{\MRDlinear}[1]{\ensuremath{\mycode{MRD}[#1]}}
\newcommand{\MRDlinq}[1]{\ensuremath{\mycode{MRD}[#1]_q^\fontmetric{R}}}
\newcommand{\Gab}[1]{\ensuremath{\mycode{G}[#1]_q^\fontmetric{R}}}

\newcommand{\codelinearHamming}[1]{\ensuremath{[#1]_q^{\fontmetric{H}}}}

\newcommand{\dimfer}{k}
\newcommand{\ferdigNoInp}{\ensuremath{\mathcal{F}}}

\newcommand{\ferdigcode}[2]{\ensuremath{[\mathcal{F},#1,#2]_q^{\fontmetric{R}}}}
\newcommand{\ferdigcodeThreeInp}[3]{\ensuremath{[#1,#2,#3]_q^{\fontmetric{R}}}}
\newcommand{\dimferdigcode}[1]{\dim(\ferdigNoInp,#1)}
\newcommand{\dimferdigcodeTwoInp}[2]{\dim(#1,#2)}
\newcommand{\numbdots}{\theta}
\newcommand{\diagonal}{\mathsf{D}}

\begin{document}

\title{Optimal Ferrers Diagram Rank-Metric Codes}
\author{\IEEEauthorblockN{Antonia Wachter-Zeh and Tuvi Etzion}\\
\IEEEauthorblockA{Computer Science Department\\ Technion---Israel Institute of Technology, Haifa, Israel\\
\texttt{\{antonia, etzion\}@cs.technion.ac.il}
\thanks{The work of Antonia Wachter-Zeh has been supported by a Minerva Postdoctoral Fellowship.}
\thanks{The work of Tuvi Etzion was supported in part by the Israeli
Science Foundation (ISF), Jerusalem, Israel, under
Grant 10/12.}
}}
\date{\today}
\maketitle

\begin{abstract}
Optimal rank-metric codes in Ferrers diagrams are considered. Such codes consist
of matrices having zeros at certain fixed positions and can
be used to construct good codes in the projective space.
Four techniques and constructions of Ferrers diagram rank-metric codes are presented,
each providing optimal codes for different diagrams and parameters.
\end{abstract}


\section{Introduction}\label{sec:introduction}
Codes in the projective space (also called \emph{subspace codes}), and in particular \emph{constant dimension codes} have become a widely investigated research topic, mostly due to their possible application to error control in random linear network coding~\cite{koetter_kschischang,SilvaKschischang-MetricsErrorCorrectionNetworkCoding_2009}.
A subspace code is a non-empty set of subspaces of a vector space of dimension $n$ over a finite field and therefore, each codeword is a subspace itself.
Constant dimension codes are special subspace codes, where each codeword has the same dimension.
The so-called \emph{subspace distance} and \emph{injection distance} are used as a distance measure for subspace codes. Several code constructions, upper bounds on the size, and properties of such codes were thoroughly investigated in~\cite{Bachoc2012Bounds,Etzion2009ErrorCorrecting,Etzion2011ErrorCorrecting,Gadouleau2010ConstantRank,KhKs09,KohnertKurz-LargeConstantDimensionCodes-2008,koetter_kschischang,silva_rank_metric_approach,Skachek2010Recursive,TrautmannManganielloRosenthal-OrbitCodes-2010,Wang2003Linear,Xia2009Johnson}.

Silva, Kschischang and K\"otter \cite{silva_rank_metric_approach} showed that lifted \emph{maximum rank distance} (MRD) codes result in subspace codes of relatively large cardinality, which can additionally be decoded efficiently in the context of random linear network coding.
MRD codes are the rank-metric analogs of Reed--Solomon codes and were introduced by Delsarte, Gabidulin and Roth \cite{Delsarte_1978,Gabidulin_TheoryOfCodes_1985,Roth_RankCodes_1991}.
Codes in the rank metric, in particular MRD codes, can be seen as a set of matrices over a finite field. The rank of the difference of two matrices is called their \emph{rank distance}, which induces a metric for such matrix codes, the \emph{rank metric}.

Lifted MRD codes \cite{silva_rank_metric_approach} provide a family of constant dimension codes. However, these codes do not attain the Singleton-like upper bound on the cardinality of constant dimension codes. Furthermore, the maximum cardinality of constant dimension codes and how to attain it is still an open question. In the same way, the maximum cardinality of codes in the projective space is not known.
Several works aim at constructing constant dimension codes or codes in the projective space of high cardinality, see e.g., \cite{Etzion2009ErrorCorrecting,EtzionSilberstein-CodesDesignsRelLiftedMRD-2012,Etzion2011ErrorCorrecting,KohnertKurz-LargeConstantDimensionCodes-2008,Skachek2010Recursive}.

The multi-level construction from \cite{Etzion2009ErrorCorrecting} is one of the constructions providing the best known cardinality for both, constant dimension codes and codes in the projective space. This is for both codes with the subspace distance and codes with the injection
distance. When the codes are constant dimension both measure distance coincide.
This construction is based on the union of several lifted rank-metric codes, which are constructed in Ferrers diagrams. The structure of the involved Ferrers diagrams is in turn defined by codewords of a constant weight code.
Informally spoken, a Ferrers diagram is an array of dots and empty entries and a Ferrers diagram rank-metric code is a set of matrices where only the entries with dots in the Ferrers diagram are allowed to be non-zero. An upper bound on the cardinality of such Ferrers diagram rank-metric codes was given in \cite{Etzion2009ErrorCorrecting}. This upper bound is a function of the diagram itself and of the desired minimum rank distance. Furthermore, a specific construction of such Ferrers diagram rank-metric codes was given in~\cite{Etzion2009ErrorCorrecting}.
The constructed codes attain the upper bound for any Ferrers diagram when the minimum rank distance is $\delta=2$.

Improvements of the multi-level construction \cite{Etzion2009ErrorCorrecting} which consider additionally so-called \emph{pending dots} of the Ferrers diagram for the construction of the subspace code were given in \cite{SilbersteinTrautmann-NewLowerBoundsForConstantDimensionCodes_2013,TrautmannRosenthal-ImprovementsEchelonFerrersConstruction_2010,SilbersteinTrautmann-SubspaceCodesBasedOnGraphmatchingFerrersDig_2014}. A dot in a Ferrers diagram is called a pending dot if the upper bound on the dimension of the rank-metric code does not change when we remove this dot.

The main goal of this paper is to construct optimal Ferrers diagram rank-metric codes, i.e., codes whose cardinality attains the upper bound from  \cite{Etzion2009ErrorCorrecting}. In this process four different techniques for
constructions of such codes are presented. These techniques have their own interest.
Each one of the four different constructions, yields good or optimal codes for different types of diagrams.
The first construction is based on \emph{maximum distance separable} (MDS) codes on the diagonals of the matrices and provides optimal codes, when---roughly spoken---the diagram has dots in the upper right triangular matrix.
The second construction can be seen as a generalization of the construction from
\cite{Etzion2009ErrorCorrecting} and is based on subcodes of MRD codes.
The third and fourth construction show how to combine two Ferrers diagram rank-metric codes
of either the same distance or the same dimension into one rank-metric code in a larger Ferrers diagram.
In the case where our constructions do not give \emph{optimal} codes, they still provide
good codes whose dimension is not far from the upper bound.
The constructed codes yield new lower bounds on the size of some subspace codes,
but this will not discuss in the paper since these small improvements
might be part of the motivation for considering Ferrers diagram rank-metric codes,
but they are not the main topic of this paper.

This paper is structured as follows.
In Section~\ref{sec:prelim}, we define Ferrers diagrams and explain known results on Ferrers diagram rank-metric codes, in particular the upper bound on the dimension of such codes.
Section~\ref{sec:construction_mds} provides our first construction based on MDS codes and Section~\ref{sec:constr_subcode} presents the second construction based on subcodes of MRD codes.
Both sections start with a toy example in order to facilitate the understanding of the constructions, then provide the constructions in a general form and finally show for which type of Ferrers diagrams \emph{optimal} rank-metric codes are obtained.
In Section~\ref{sec:combination_diagrams}, we show two methods for combining existing Ferrers diagram rank-metric codes to obtain codes in larger diagrams.
Section~\ref{sec:comparison_andthree} compares the different constructions
and finally, in Section~\ref{sec:conclusion}, we conclude the paper and present a few open problems.

\section{Ferrers Diagrams and Ferrers Diagram Rank-Metric Codes}\label{sec:prelim}
\subsection{Notations}
Let $q$ be a power of a prime and let $\Fq$ denote the finite field of order $q$ and $\Fqm$ its extension field of order~$q^m$.
We use $\Fq^{s \times n}$ to denote the set of all $s\times n$ matrices over $\Fq$ and $\Fqm^n$ to denote the set of all row vectors of length~$n$ over $\Fqm$.
Let $\mathbf I_{s}$ denote the $s \times s$ identity matrix, $\intervallincl{a}{b}$ the set of integers $\{i: a \leq i \leq b, i \in \mathbb{Z}\}$, and let the rows and columns of an $m\times n$ matrix $\mathbf A$ be indexed by $0,\dots, m-1$ and $0,\dots, n-1$, respectively.
Throughout this paper, we assume w.l.o.g. that $n \leq m$.

\subsection{Rank-Metric Codes}

We define the \textit{rank norm} $\rk(\A)$ as the rank of $\A\in \Fq^{m \times n}$ over $\mathbb{F}_{q}$.
The \emph{rank distance} between $\A \in \Fq^{m\times n}$ and $\B \in \Fq^{m\times n}$ is the rank of the difference of the two matrix representations (compare \cite{Gabidulin_TheoryOfCodes_1985}):
\begin{equation*}
d_{\fontmetric{R}}(\A,\B)\defeq  \rk(\A-\B).
\end{equation*}
An $\codelinear{m \times n,k,\delta}$ \emph{rank-metric code} \mycodeRank{C} denotes a linear rank-metric code, i.e.,
it is a $k$-dimensional linear subspace of $\Fq^{m \times n}$. It consists of $M = q^k$ matrices of size $m \times n$ over $\Fq$ and has minimum rank distance $\delta$, which is defined by
\begin{equation*}
\delta \defeq \min_{\substack{{\A,\B} \in \mycodeRank{C}\\ \A \neq \B}}
\big\lbrace d_{\fontmetric{R}}(\A,\B) =
\rk(\A-\B) \big\rbrace.
\end{equation*}
The Singleton-like upper bound for rank-metric codes \cite{Delsarte_1978,Gabidulin_TheoryOfCodes_1985,Roth_RankCodes_1991} implies that for any $\codelinear{m \times n,k,\delta}$ code, the dimension is bounded from above by $k \leq \max\{m,n\}(\min\{n,m\}-\delta+1)$.
If a code attains this bound with equality, it is called a \emph{maximum rank distance} (MRD) code.
The notation $\MRDlinq{m\times n, \delta}$ denotes a linear MRD code (in matrix representation), consisting of matrices of size $m\times n$ over $\Fq$ with minimum rank distance $\delta$. 

\subsection{Ferrers Diagrams and Rank-Metric Codes in Ferrers Diagrams}
\emph{Ferrers diagrams} are used to represent partitions by arrays of dots and empty entries and are defined as follows (see also~\cite{AndrewsEriksson-IntegerPartitions_Book} and compare Example~\ref{ex:ferrers_diagram}).

\begin{example}\label{ex:ferrers_diagram}
The following example shows a $5 \times 4$ Ferrers diagram $\ferdigNoInp$.
\begin{equation*}
\ferdigNoInp = \quad \begin{matrix}
\bullet & \bullet & \bullet & \bullet \\
\bullet & \bullet & \bullet & \bullet \\
&  & \bullet & \bullet \\
&  & \bullet & \bullet \\
&&&\bullet
\end{matrix}.
\end{equation*}
\end{example}

\begin{definition}[Ferrers Diagram]
An $m \times n$ Ferrers diagram $\ferdigNoInp$, where $n \leq m$, is an $m \times n$ array of dots and empty entries with the following properties:
\begin{itemize}
\item all dots are shifted to the right,
\item the number of dots in each row is at most the number of dots in the previous row,
\item the first row has $n$ dots and the rightmost column has $m$ dots.
\end{itemize}
\end{definition}

Motivated by the multi-level construction from \cite{Etzion2009ErrorCorrecting}, we want to construct good or even optimal rank-metric codes in Ferrers diagrams, i.e., a set of matrices having non-zero elements only at positions where the Ferrers diagram has dots and with a certain minimum rank distance between the different matrices.

For a given $m \times n$ Ferrers diagram $\ferdigNoInp$, the triple $\ferdigcode{\dimfer}{\delta}$ denotes a linear Ferrers diagram rank-metric code over~$\Fq$ of minimum rank distance $\delta$. The code consists of matrices in $\Fq^{m \times n}$, the dimension of this code is  $\dimfer$ and its cardinality therefore $q^\dimfer$. The code is defined such that the entries of all matrices are zero where the corresponding Ferrers diagram has no dots. For given $\ferdigNoInp$ and $\delta$, the maximum dimension of an associated rank-metric code is denoted by $\dimferdigcode{\delta}$ and clearly, $\dimfer \leq \dimferdigcode{\delta}$.


Further, by $\gamma_i$, $\forall i \in \intervallexcl{0}{n}$, we denote the number of dots in $\ferdigNoInp$ in the $i$-th column and by $\rho_i$, $\forall i \in \intervallexcl{0}{m}$, the number of dots in $\ferdigNoInp$ in the $i$-th row.

\subsection{Known Results and Problem Statement}
An upper bound on $\dimferdigcode{\delta}$ was given in \cite[Theorem~1]{Etzion2009ErrorCorrecting} and a construction achieving this bound for specific diagrams (compare Theorem~\ref{thm:es-construction}) was shown.

\begin{theorem}[Upper Bound {\cite[Theorem~1]{Etzion2009ErrorCorrecting}}]\label{thm:upper_bound}
Let $\nu_i$, $i \in \intervallexcl{0}{\delta}$, denote the number of dots in a Ferrers diagram $\ferdigNoInp$ after removing the first $i$ rows and the $\delta-1-i$ rightmost columns. Then,
\begin{equation*}
\dimferdigcode{\delta} \leq \min_{i \in \intervallexcl{0}{\delta}}\nu_i.
\end{equation*}
\end{theorem}
Throughout this paper, codes which attain this bound are called \emph{optimal}.
The following problem will be investigated in this paper.
\begin{problem}[Optimal Ferrers Diagram Rank-Metric Codes]
Let $\ferdigNoInp$ be an $m \times n$ Ferrers diagram, $q \geq 2$ be a prime power, and $\delta$ be a positive integer, where $\delta \leq n \leq m $.
Does there exist an $\ferdigcode{\dimfer}{\delta}$ rank-metric code $\mycodeRank{C}$ which attains the upper bound from Theorem~\ref{thm:upper_bound}?
\end{problem}
We want to find general code constructions, which provide optimal Ferrers diagram rank-metric codes for large classes of diagrams and many values of $\delta$.

\begin{theorem}[Construction from {\cite{Etzion2009ErrorCorrecting}}]\label{thm:es-construction}
Let $\ferdigNoInp$ be an $m \times n$ Ferrers diagram and assume that each of the $\delta-1$ rightmost columns has $m$ dots. Then, the construction from \cite{Etzion2009ErrorCorrecting} provides an $\ferdigcode{\dimfer}{\delta}$ rank-metric code attaining the bound from Theorem~\ref{thm:upper_bound} for any $q \geq 2$ and any $\delta$, and therefore, $\dimfer = \min_{i \in \intervallexcl{0}{\delta}}\nu_i$.
\end{theorem}
In particular, this construction gives optimal codes for \emph{any} diagram when $\delta=2$ since we clearly always have $\delta-1 = 1$ columns with $m$ dots. The construction was described in \cite{Etzion2009ErrorCorrecting} by means of $q$-cyclic MRD codes, but
it can also be interpreted using systematic shortened MRD codes, see also \cite{GabidulinPilipchuk-RankSubcodesMulticomponentNetworkCoding_2013,KhaleghiSilvaKschischang-SubspaceCodes_2009}.

By shortening systematic MRD codes, we mean that a systematic $m \times n$ MRD code of dimension $k = m(n-\delta+1)$ is constructed and afterwards all matrices with non-zero entries at the empty entries in the Ferrers diagram are discarded. If empty entries occur only at the \emph{information} positions of the MRD code, the shortened code will be an optimal Ferrers diagram rank-metric code. However, if we have such positions in the \emph{redundancy} part (which can be chosen to be the $\delta-1$ rightmost columns), then this construction will not be optimal.

\section{Construction based on MDS Codes}\label{sec:construction_mds}

In this section, we present a construction of Ferrers diagram rank-metric codes based on \emph{maximum distance separable} (MDS) codes.

In the sequel, the triple $\codelinearHamming{n,k,d}$ denotes a linear code over $\Fq$ of length $n$, dimension $k$, and minimum Hamming distance $d$.
The Singleton upper bound states that for any $\codelinearHamming{n,k,d}$ code, $d \leq n-k+1$ holds. Codes which attain the Singleton bound are called MDS codes.
An  $\codelinearHamming{n,k=n-d+1,d}$ MDS code exists if the field size is sufficiently large, i.e., $q \geq n-1$ or when $d\in \{1,2,n \}$, see \cite[Ch.~11, Thm.~9]{MacWilliamsSloane_TheTheoryOfErrorCorrecting_1988}.

Our construction based on MDS codes will first be described by an example.
For this construction, we need the notation of the \emph{Hamming weight} of a vector $\vec{a} \in \Fq^n$, which is $\wtH{\vec{a}}$.
\begin{example}\label{ex:fourtimesfour_triang}
Let $\ferdigNoInp$ be the following $5 \times 5$ Ferrers diagram:
\begin{equation*}
\begin{matrix}
\bullet & \bullet & \bullet & \bullet& \bullet\\
& \bullet & \bullet & \bullet& \bullet\\
&& \bullet & \bullet& \bullet\\
&&& \bullet& \bullet\\
&&&& \bullet
\end{matrix}.
\end{equation*}
For $\delta=2$, an optimal code for $\ferdigNoInp$ was given in~\cite{Etzion2009ErrorCorrecting}.
We will show a general construction and use this example with $\delta=3$ only to introduce the idea. The upper bound from Theorem~\ref{thm:upper_bound} for $\delta = 3$ is $\dimferdigcodeTwoInp{\ferdigNoInp}{\delta} \leq 6$.

Let $q \geq 4$, let $\mycode{A}_3$ be a $\codelinearHamming{3,1,3}$ MDS code, $\mycode{A}_4$ a $\codelinearHamming{4,2,3}$ MDS code, and $\mycode{A}_5$ a $\codelinearHamming{5,3,3}$ MDS code.
Let $\vec{a}=(a_0 \ a_1 \ a_2)$, $\vec{b}=(b_0 \ b_1 \ b_2\ b_3)$, and $\vec{c}=(c_0 \ c_1 \ c_2\ c_3 \ c_4)$ and define the following rank-metric code $\mycodeRank{C}$:
\begin{equation*}
\mycodeRank{C}
= \left\lbrace
\begin{pmatrix}
c_0 & b_0& a_0 & 0 & 0\\
0& c_1 & b_1&a_1 & 0\\
0&0& c_2 & b_2& a_2\\
0&0&0& c_3& b_3\\
0&0&0&0 & c_4\\
\end{pmatrix}:
\vec{a}
\in \mycode{A}_3,
\vec{b}
\in \mycode{A}_4,
\vec{c}
\in \mycode{A}_5
\right\rbrace.
\end{equation*}
The assumption $q \geq 4$ guarantees that the MDS codes $\mycode{A}_3$, $\mycode{A}_4$, and $\mycode{A}_5$ exist.
The cardinality of $\mycodeRank{C}$ is clearly $|\mycode{A}_3|\cdot |\mycode{A}_4|\cdot |\mycode{A}_5|=q^6$ and therefore its dimension is $\dimfer=6$, attaining the upper bound from Theorem~\ref{thm:upper_bound}.

Furthermore, $\mycodeRank{C}\subseteq \Fq^{5 \times 5}$ is linear in~$\Fq$, since $\mycode{A}_3, \mycode{A}_4, \mycode{A}_5$ are linear codes. Hence, for calculating the minimum rank distance of $\mycodeRank{C}$, it is sufficient to calculate the minimum rank weight.
For this purpose, notice that
\begin{equation*}
\rk\left(\begin{matrix}
\Mat{A} & \Mat{B}\\
\0 & \Mat{D}\\
\end{matrix}\right) \geq \rk(\Mat{A})+\rk(\Mat{D}),
\end{equation*}
for any three matrices $\Mat{A}, \Mat{B}, \Mat{D}$ of suitable sizes. We can apply this fact recursively on any codeword in $\mycodeRank{C}$, leading to the following considerations.
Let $\C$ be a codeword of $\mycodeRank{C} $. If $\vec{c}\neq \0$, then $\rk(\C) \geq \wtH{\c} \geq 3$; if $\c =\0$ and $\b \neq \0$, then $\rk(\C)\geq \wtH{\b} \geq 3$; and if $\c =\b=\0$ and $\a \neq \0$, then $\rk(\C) = \wtH{\a} = 3$.
Therefore, the minimum rank weight of any non-zero codeword of $\mycodeRank{C}$ is three and therefore $\delta = 3$.

Thus, $\mycodeRank{C}$ is an optimal $\ferdigcode{6}{3}$ rank-metric code for any $q \geq 4$.
\end{example}
Notice that a similar idea with MDS codes on the diagonals was used in \cite[Section~V]{Roth_RankCodes_1991} to construct rank-metric codes over infinite fields and codes in the so-called cover metric for correcting crisscross errors.

Example~\ref{ex:constr_mds_one} is generalized in Construction~\ref{def:mds_construction}, where we refer to \emph{diagonals} of a Ferrers diagram as follows.

\begin{definition}[Diagonals of Ferrers Diagram]
\label{def:diagonals_ferrers}
A diagonal of a Ferrers diagram $\ferdigNoInp$ is a consecutive sequence of entries, going upwards diagonally from the rightmost column to either the leftmost column or the first row.
Let $\diagonal_i$, $\forall i\in\intervallincl{0}{m-1}$, denote the $i$-th diagonal in $\ferdigNoInp$, where $i$ counts the diagonals from the top to the bottom and let $\numbdots_i$ denote the number of dots on $\diagonal_i$ in $\ferdigNoInp$.
\end{definition}
Note that in the rightmost column there must be a dot in each diagonal and therefore $1 \leq \numbdots_i \leq n$, $\forall i \in \intervallexcl{0}{m}$.

To each $m\times n$ Ferrers diagram, we associate a set of matrices (i.e., our codewords) and therefore, when we speak about a diagonal, we refer to both, the diagonals in the Ferrers diagram as well as the corresponding diagonals in the associated matrices.
For an $m\times n$ diagram, we can illustrate the $m$ diagonals as follows:

\begin{center}
\includegraphics[scale=1.05]{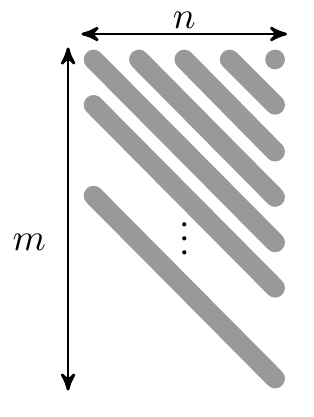}
\end{center}

\begin{construction}[Based on MDS Codes]
\label{def:mds_construction}
Let $\ferdigNoInp$ be an $m \times n$ Ferrers diagram and let $\delta$ be an integer such that $0 <\delta \leq n$.
Let $\numbdots_{max} = \max_{{i\in \intervallexcl{0}{m}}} \numbdots_i$ and let $q \geq \numbdots_{max}-1$.

Let $\mycode{A}_j$ be a $\codelinearHamming{j,j-\delta+1,\delta}$ MDS code, for all $j \in \intervallincl{\delta}{\numbdots_{max}}$, and let $\mycode{A}_j = \{ \0 \}$ for all $j \in \intervallexcl{0}{\delta}$.
Define the following rank-metric code $\mycodeRank{C}$:
\begin{equation*}
\mycodeRank{C}
= \left\lbrace
\Mat{C} \in \Fq^{m \times n} \ : \
\diagonal_i \in \mycode{A}_{\numbdots_i}, \forall i \in \intervallexcl{0}{m}
\right\rbrace.
\end{equation*}
\end{construction}
Note that by $\diagonal_i \in \mycode{A}_{\numbdots_i}$, we actually refer only to those positions on $\diagonal_i$ where $\ferdigNoInp$ has dots.

\begin{lemma}[Parameter of Codes]
Construction~\ref{def:mds_construction} provides a linear $\ferdigcode{\dimfer}{\delta}$ rank-metric code~$\mycodeRank{C}$ over~$\Fq$ of dimension $\dimfer = \sum_{i=0}^{m-1} \min\{0, \numbdots_i-\delta+1\}$ for any $q \geq \numbdots_{max}-1$.
\end{lemma}
\begin{IEEEproof}
The assumption $q \geq \numbdots_{max}-1$ guarantees that (linear) MDS codes of length at most $\numbdots_{max}$ over $\Fq$ exist.
The code $\mycodeRank{C}$ is linear, since all the MDS codes are linear.

The cardinality of $\mycodeRank{C}$ is the multiplication of the cardinalities of the MDS codes and therefore, the dimension of~$\mycodeRank{C}$ is $\dimfer=\sum_{i=01}^{m-1}\min\{0, \numbdots_i-\delta+1\}$.

Since $\mycodeRank{C}$ is linear, to calculate $\delta$, it is sufficient to calculate the minimum rank \emph{weight} of any non-zero codeword.
Recall that
\begin{equation*}
\rk\left(\begin{matrix}
\Mat{A} & \Mat{B}\\
\0 & \Mat{D}\\
\end{matrix}\right) \geq \rk(\Mat{A})+\rk(\Mat{D}),
\end{equation*}
holds for any three matrices $\Mat{A}, \Mat{B}, \Mat{D}$ of suitable sizes. To apply this fact, we can cut any code matrix recursively into smaller matrices and therefore, the rank of the code matrix is at least the Hamming weight of the bottommost non-zero diagonal. Since each non-zero diagonal is a codeword of an MDS code of minimum Hamming weight $\delta$, the rank of any non-zero codeword is at least $\delta$.
\end{IEEEproof}

When $q$ is sufficiently large, we can therefore construct Ferrers diagram rank-metric codes (which are not necessarily optimal) for any diagram with this strategy.
The following theorem shows that for many Ferrers diagrams, this construction provides \emph{optimal} rank-metric codes.

\begin{figure}[htb, scale = 1]
\centering
\includegraphics[scale=1]{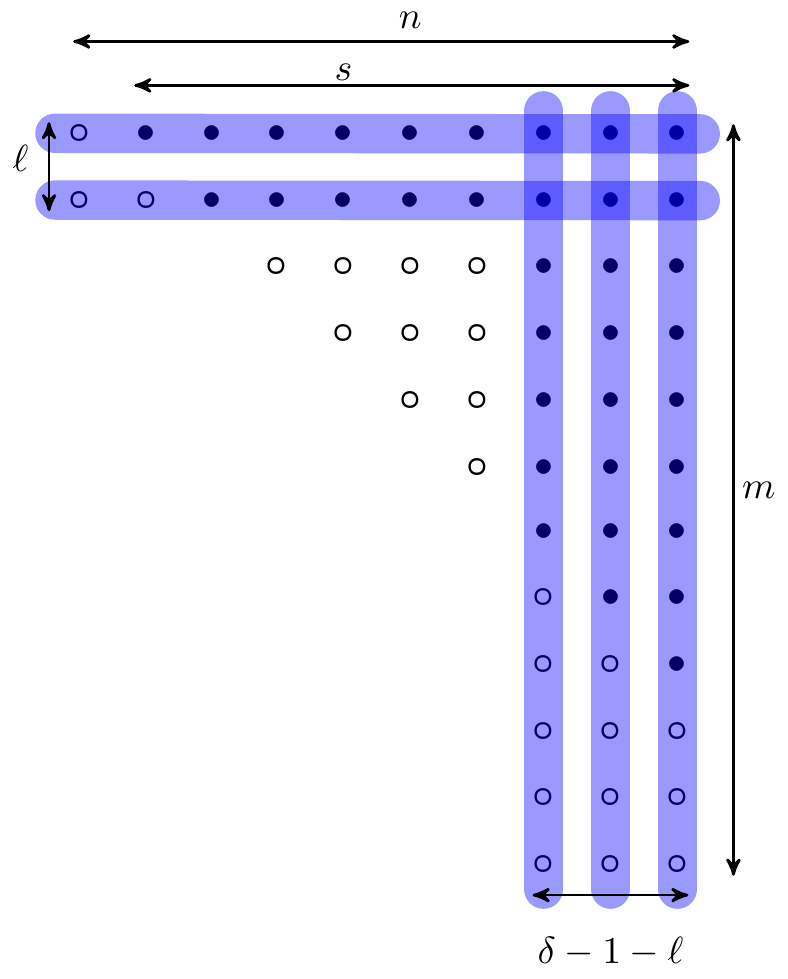}

\caption{Illustration of the Ferrers diagrams, for which Construction~\ref{def:mds_construction} yields optimal codes. The dots "$\bullet$" in the blue shaded area have to exist, whereas the dots marked by "$\boldsymbol{\circ}$" can exist or not. 
\label{fig:ferrdig_assumptheorem_mds}}
\hrulefill
\end{figure}

\begin{theorem}[Optimality of Construction~\ref{def:mds_construction}]\label{thm:optimality_mds_constr}
Let $\ferdigNoInp$ be an $m \times n$ Ferrers diagram and let $\delta$ be an integer, $0 < \delta \leq n$.
Assume that $\ell$ rows and $\delta-1-\ell$ columns have to be removed to obtain the upper bound on $\dimferdigcode{\delta}$ from Theorem~\ref{thm:upper_bound}.

Further, assume that there is some $s \in \intervallexcl{0}{m}$ such that are no dots below the diagonal $\diagonal_{s-1}$, apart from dots in the first $\ell$ rows and $\delta-1-\ell$ rightmost columns (see also Figure~\ref{fig:ferrdig_assumptheorem_mds}), and that there are $\delta-1$ dots on $\diagonal_{s-1}$ which are removed for the upper bound.

Then, for any $q \geq \numbdots_{max}-1$, Construction~\ref{def:mds_construction} provides optimal Ferrers diagram rank-metric codes, i.e., their dimension attains the upper bound from Theorem~\ref{thm:upper_bound}. 
\end{theorem}

\begin{IEEEproof}
Consider a complete triangular $s \times s$ Ferrers diagram $\widehat{\ferdigNoInp}$, depicted below:
\begin{equation*}
\begin{matrix}
\bullet & \bullet&\bullet & \dots & \bullet & \bullet& \bullet\\
&\bullet&\bullet &  & \bullet & \bullet& \bullet\\
&& \bullet& \dots & \bullet& \bullet& \bullet \\
&&&\ddots&\vdots  &&\vdots\\
&&& & \bullet & \bullet & \bullet\\
&&&& &\bullet & \bullet\\
&&&&& &\bullet\\
\end{matrix}
\end{equation*}
For such a symmetric diagram, it does not matter if we delete rows or columns for the calculation of the upper bound since
\begin{equation*}
\dimferdigcodeTwoInp{\widehat{\ferdigNoInp}}{\delta} \leq\nu_0=\nu_1=\dots=\nu_{\delta-1} = \sum_{i=1}^{s-\delta+1}i = \frac{(s-\delta+1)(s-\delta+2)}{2},
\end{equation*}
where $\nu_i$, $\forall i\in \intervallexcl{0}{\delta}$, is defined as in Theorem~\ref{thm:upper_bound}.

Construction~\ref{def:mds_construction} places an $\codelinearHamming{s,s-\delta+1,\delta}$  MDS code on $\diagonal_{s-1}$ of $\widehat{\ferdigNoInp}$ and $\codelinearHamming{j,j-\delta+1,\delta}$ MDS codes on the diagonals $\diagonal_{j-1}$, $\forall j\in \intervallincl{\delta}{s-1}$. Therefore, $\widehat{\dimfer} = \sum_{i=1}^{s-\delta+1}i$, which attains the bound from Theorem~\ref{thm:upper_bound}.

Furthermore, it is easy to verify that deleting $\Delta$ dots, which are not in the first $\ell$ rows or the $\delta-1-\ell$ rightmost columns, decreases both, the upper bound as well as the dimension of the constructed code by $\Delta$.


%

Finally, adding "pending dots", i.e., additional dots in the first $\ell$ rows and the $\delta-1-\ell$ rightmost columns, changes neither the upper bound nor the dimension of the constructed code and Construction~\ref{def:mds_construction} gives optimal codes.
\end{IEEEproof}

Notice that there are diagrams, where $\ell$ can be chosen in several ways.


For $\ell=0$, Theorem~\ref{thm:optimality_mds_constr} includes some cases covered by the construction of \cite{Etzion2009ErrorCorrecting}. We require a larger field size than the construction of \cite{Etzion2009ErrorCorrecting}, but for many diagrams, our construction gives optimal Ferrers diagram rank-metric codes, where the construction from \cite{Etzion2009ErrorCorrecting} does not.

Let us show a few examples of Ferrers diagrams, where Construction~\ref{def:mds_construction} yields optimal codes.

\begin{example}\label{ex:constr_mds_one}
For $\ell=0$, $s=m -1 $, $q \geq n-1$, Construction~\ref{def:mds_construction} provides optimal codes for the following class of $m \times n$ diagrams:
\begin{equation*}
\begin{matrix}
\bullet &\bullet & \bullet & \dots &\bullet & \bullet\\
\bullet &\bullet & \bullet & \dots &\bullet & \bullet\\
\bullet &\bullet & \bullet & \dots &\bullet & \bullet\\
\bullet &\bullet & \bullet & \dots &\bullet & \bullet\\
&\bullet & \bullet & \dots &\bullet & \bullet\\
&& \bullet & \dots &\bullet & \bullet\\
&&&\ddots& \ddots &\vdots \\
&&&& \bullet& \bullet\\
&&&& & \bullet\\
\end{matrix}.
\end{equation*}
\end{example}
\begin{example}
Consider the following $8 \times 5$ Ferrers diagram and let $\delta=3$.
The upper bound from Theorem~\ref{thm:upper_bound} is $\dimferdigcode{{\delta}} \leq \sum_{i=0}^{2} \gamma_i = 10$.
This diagram belongs to the class of diagrams of Theorem~\ref{thm:optimality_mds_constr} with $\ell=0$, $s=m-2=6$ and $q \geq 4$.
\begin{equation*}
\begin{matrix}
\bullet &\bullet & \bullet  &\bullet & \bullet\\
\bullet &\bullet & \bullet  &\bullet & \bullet\\
 &\bullet & \bullet  &\bullet & \bullet\\
 &\bullet & \bullet  &\bullet & \bullet\\
& &  &\bullet & \bullet\\
&&  &\bullet & \bullet\\
&&&& \bullet& \\
&&&&  \bullet\\
\end{matrix}.
\end{equation*}
On $\diagonal_4$, we place an $\codelinearHamming{5,3,3}$ code, the same on $\diagonal_5$. Further, on $\diagonal_3$, we place an $\codelinearHamming{4,2,3}$ code and on $\diagonal_2$ an $\codelinearHamming{3,1,3}$ code. On $\diagonal_6$, we also place an $\codelinearHamming{3,1,3}$ code. Hence, we attain the upper bound, since $\dimfer = 3+3+2+1+1=10$.
\end{example}

\begin{example}
Consider the following $6 \times 6$ Ferrers diagram, let $\delta=3$.
The upper bound from Theorem~\ref{thm:upper_bound} gives $\dimferdigcode{{\delta}} \leq 5$ when $\ell=1$ rows and $\delta-1-\ell=1$ columns are removed.
This diagram belongs to the class of diagrams of Theorem~\ref{thm:optimality_mds_constr} with $\ell=1$, $s=4$, and $q \geq 3$.
\begin{equation*}
\begin{matrix}
\bullet &\bullet &\bullet & \bullet  &\bullet & \bullet\\
 && & \bullet  &\bullet & \bullet\\
 && & \bullet  &\bullet & \bullet\\
 && &  &\bullet & \bullet\\
& &&  & & \bullet\\
&& & & & \bullet\\
\end{matrix}.
\end{equation*}
We place an $\codelinearHamming{3,1,3}$ MDS code on $\diagonal_2$, an $\codelinearHamming{4,2,4}$ MDS code on $\diagonal_3$ and an $\codelinearHamming{4,2,4}$ MDS code on $\diagonal_4$. Thus, $k=5$, attaining the upper bound from Theorem~\ref{thm:upper_bound}.
\end{example}

\section{Construction based on Subcodes of MRD Codes}\label{sec:constr_subcode}
\subsection{Matrix/Vector Representation and Gabidulin Codes}
Let $\NormbasisOrdered= \vecelementsm{\beta}$ be an ordered basis of $\Fqm$ over $\Fq$. There is a bijective map $\extsmallfielddeg{m}$ of any vector $\mathbf a \in \Fqm^n$ on a matrix $\mathbf A \in \Fq^{m \times n}$, denoted as follows:
\begin{align*}
\extsmallfielddeg{m}:\ \Fqm^{n} &\rightarrow \Fq^{m \times n}\label{eq:mapping_smallfield}\\
\a = \vecelements{a} &\mapsto \A,
\end{align*}
where $\A =\extsmallfieldinputdeg{\vec{a}}{m}\in \Fq^{m \times n}$ is defined such that
\begin{equation*}
\quad a_j = \sum_{i=0}^{m-1} A_{i,j} \Normelement_i, \quad \forall j \in \intervallexcl{0}{n}.
\end{equation*}
The map $\extsmallfielddeg{m}$ will be used to facilitate switching between a vector in $\Fqm$ and its matrix representation over $\Fq$.
In the sequel, we use both representations, depending on what is more convenient in the context and by slight abuse of notation, $\rk(\a)$ denotes $\rk(\extsmallfieldinputdeg{\a}{m})$.

Gabidulin codes are a special class of MRD codes and the vector representation of its codewords can be defined by a generator matrix as follows, where we denote the $q$-power for any positive integer $i$ and any $a \in \Fqm$ by $a^{[i]}\defeq a^{q^i}$.

\begin{definition}[Gabidulin Code \cite{Gabidulin_TheoryOfCodes_1985}]
A linear $\Gab{m \times n,\delta}$ MRD code, in vector representation over $\Fqm^n$, of dimension $m(n-\delta+1)$ and minimum rank distance $\delta$ is defined by its $(n-\delta+1) \times n$ generator matrix $\mathbf G$:
\begin{equation*}
\mathbf G =
\MoormatExplicitIncl{g}{n-\delta}{n},
\end{equation*}
where $\extsmallfieldinputdeg{g_0}{m}, \extsmallfieldinputdeg{g_1}{m}, \dots, \extsmallfieldinputdeg{g_{n-1}}{m} \in \Fq^m$ are linearly independent over $\Fq$.
\end{definition}
As mentioned before, a $\Gab{m \times n,\delta}$ code is an $\MRDlinq{m\times n, \delta}$ code, see \cite{Gabidulin_TheoryOfCodes_1985}.

\subsection{The Subcode Construction}
Our second construction is based on a subcode of a systematic MRD code, and it provides optimal codes for several diagrams, for which previously no optimal codes were known.
We first illustrate the idea with an example. Notice that for this example, neither the construction from \cite{Etzion2009ErrorCorrecting} nor Construction~\ref{def:mds_construction} give optimal codes.
\begin{example}\label{ex:fourtimesfour_nontriang}
Let $\ferdigNoInp$ be the following $4\times 4$ Ferrers diagram:
\begin{equation*}
\ferdigNoInp=\quad
\begin{matrix}
\bullet & \bullet & \bullet & \bullet\\
& \bullet & \bullet & \bullet\\
&\bullet& \bullet & \bullet\\
&&& \bullet\\
\end{matrix}.
\end{equation*}
For $\delta=3$, the upper bound gives $\dimferdigcode{\delta} \leq 4$.
The construction from \cite{Etzion2009ErrorCorrecting} does not give optimal codes here, since there are no two columns (or two rows) with $m=4$ dots.
Similarly, Construction~\ref{def:mds_construction} gives only $\dimfer=3$ for $q \geq 3$.
In the following, we show the idea of a (general) construction which provides optimal codes for this diagram.

Let
\begin{equation}\label{eq:gemat_gab_example}
\Mat{A} \cdot
\underbrace{\begin{pmatrix}
g_0 & g_1 & g_2 \\
g_0^{[1]} & g_1^{[1]}  & g_2^{[1]}   \\
\end{pmatrix}}_{\defeq \Mat{G}}
=
\begin{pmatrix}
1 & 0 & \alpha_{0,2}\\
0 & 1 & \alpha_{1,2}\\
\end{pmatrix}
\in \Fqext{3}^{2 \times 3},
\end{equation}
be a systematic generator matrix of an $\Gab{3 \times 3, 2}$ MRD code (i.e., $\Mat{A}$ is a full-rank matrix in $\Fqext{3}^{2 \times 2}$ and ${g_0,g_1,g_2 \in \Fqext{3}}$ are linearly independent over $\Fq$).
We denote this $\Gab{3 \times 3, 2}$ MRD code by $\mycodeRank{A}$.

Further, let
\begin{equation*}
\widehat{\Mat{G}} \defeq
\begin{pmatrix}
1 & 0 & \alpha_{0,2}& 0\\
0 & 1 & \alpha_{1,2}& \alpha_{1,3}\\
\end{pmatrix}.
\end{equation*}
Since the vector $(0 \ 1 \ \alpha_{1,2}) \in \Fqext{3}^3$ is a linear combination of the rows of $\Mat{G}$, it is a codeword of $\mycodeRank{A}$ and hence, it has $\rk(\extsmallfieldinputdeg{1 \ \alpha_{1,2}}{m}) = 2$. Additionally, we choose $\alpha_{1,3} \in \Fqext{3}$ such that $\rk(\extsmallfieldinputdeg{1 \ \alpha_{1,2}\ \alpha_{1,3}}{m}) = 3$.

The set
\begin{equation*}
\mycodeRank{\widehat{C}} = \left\lbrace\vec{u} \cdot \widehat{\Mat{G}}\ : \ \vec{u} = (u_0 \ u_1)\in \Fqext{3}^2, \ \extsmallfieldinputdeg{u_0}{3} =
\begin{pmatrix}
u_{0,0}\\
0\\
0
\end{pmatrix}\!,
\ \extsmallfieldinputdeg{u_1}{3} =
\begin{pmatrix}
u_{1,0}\\
u_{1,1}\\
u_{1,2}\\
\end{pmatrix}\!,
\ u_{0,0},u_{1,0},u_{1,1}u_{1,2} \in \Fq\right\rbrace
\end{equation*}
is a code of cardinality $q^4$.
The first three symbols $(c_0 \ c_1 \ c_2)$ of each codeword $\c= (c_0 \ c_1 \ c_2\ c_3)\in \mycodeRank{\widehat{C}} \subseteq \Fqext{3}^4$ form a codeword of the systematic MRD code $\mycodeRank{A}$. The matrix representation of a codeword $\vec{c} \in \mycodeRank{\widehat{C}}$ is:
\begin{equation*}
\extsmallfieldinputdeg{\vec{c}}{3} =
\begin{pmatrix}
u_{0,0} & u_{1,0}  & c_{2,0} & c_{3,0}\\
0& u_{1,1}  & c_{2,1} &  c_{3,1}\\
0&u_{1,2} & c_{2,2} &  c_{3,2}\\
\end{pmatrix}.
\end{equation*}
We associate the matrix representation of each codeword of $\mycodeRank{\widehat{C}}$ with the first three rows of the Ferrers diagram $\ferdigNoInp$ and additionally, on the bottom right corner, we place the repetition of $u_{0,0}$. Let $\mycodeRank{C}$ be the set of all such matrices, i.e.:
\begin{equation*}
\mycodeRank{C} =
\left\lbrace
\begin{pmatrix}
u_{0,0} & u_{1,0}  & c_{2,0} & c_{3,0}\\
0& u_{1,1}  & c_{2,1} &  c_{3,1}\\
0&u_{1,2} & c_{2,2} &  c_{3,2}\\
0&0&0& u_{0,0} \\
\end{pmatrix}
\right\rbrace.
\end{equation*}
As explained before, $\mycodeRank{C}$ is an $\ferdigcode{\dimfer}{\delta}$ code of dimension $\dimfer = 4$ and it is clearly linear. It remains to show the minimum rank weight of any non-zero codeword of $\mycodeRank{C}$.
We distinguish between two cases:
\begin{itemize}
\item $u_{0,0} \neq 0$. The matrix
\begin{equation*}
C \defeq
\begin{pmatrix}
u_{0,0} & u_{1,0}  & c_{2,0}\\
0& u_{1,1}  & c_{2,1}\\
0&u_{1,2} & c_{2,2} \\
\end{pmatrix}
\end{equation*}
is a codeword of $\mycodeRank{A}$ and has therefore rank at least two. Hence, the overall rank of a non-zero codeword of $\mycodeRank{C}$ is $\rk(C)+\rk(u_{0,0}) \geq 3$.
\item $u_{0,0} = 0$. Clearly, $\vec{c} = u_1 \cdot (0 \ 1 \ \alpha_{1,2}\ \alpha_{1,3}) \in \mycodeRank{\widehat{C}}$.
Since $\rk(\extsmallfieldinputdeg{1 \ \alpha_{1,2}\ \alpha_{1,3}}{m}) = 3$, it follows that $(1 \ \alpha_{1,2}\ \alpha_{1,3})$ is a generator matrix of an $\MRDlinq{3 \times 3,3}$ code and the rank of $\extsmallfieldinputdeg{\vec{c}}{3}$ is therefore three.
\end{itemize}
Thus, this construction provides an optimal $\ferdigcode{4}{3}$ code, for any $q \geq 2$.
\end{example}

Notice further that the same strategy also gives optimal Ferrers diagram rank-metric codes for the diagram of Example~\ref{ex:fourtimesfour_triang} and any larger triangular diagram if $\delta=3$. However, for higher $\delta$, the construction based on subcodes of MRD codes does not provide optimal codes for triangular diagrams (compare also Theorem~\ref{thm:optimality_mrd_subcodes} about the optimality of the subcode construction), whereas Construction~\ref{def:mds_construction} does when $q$ is sufficiently large.
Therefore, there are also diagrams for which Construction~\ref{def:mds_construction} provides optimal codes, but the second construction does not.

Let us now generalize the construction from Example~\ref{ex:fourtimesfour_nontriang}. For this purpose, we need the following lemma, whose proof can be found in the appendix.

\begin{lemma}[Systematic Generator Matrices]\label{lem:syst_gen_mat}
For $\eta-1\leq \mu$, there exists a $\kappa \times \eta$ matrix of the following form:
\begin{equation*}
\Mat{G}_{\kappa \times \eta} =
\begin{pmatrix}
1&&& &\alpha_{\kappa,0}&\dots & \alpha_{\eta-2,0} &0\\
&1&&& \alpha_{\kappa,1}&\dots & \alpha_{\eta-2,1} &\alpha_{\eta-1,1}\\
&&\ddots&& \vdots&& \vdots\\
&&&1& \alpha_{\kappa,\kappa-1}&\dots & \alpha_{\eta-2,\kappa-1}& \alpha_{\eta-1,\kappa-1}\\
\end{pmatrix}
\in \Fqext{\mu}^{\kappa \times \eta},
\end{equation*}
such that the $\kappa \times (\eta-1)$ submatrix
\begin{equation*}
\Mat{G}_{\kappa\times (\eta-1)} \defeq
\begin{pmatrix}
1 &&& &\alpha_{\kappa,0}&\dots & \alpha_{\eta-2,0}\\
&1&&& \alpha_{\kappa,1}&\dots & \alpha_{\eta-2,1} \\
&&\ddots&& \vdots&\ddots& \vdots\\
&&&1& \alpha_{\kappa,\kappa-1}&\dots & \alpha_{\eta-2,\kappa-1}\\
\end{pmatrix}
\end{equation*}
is a systematic generator matrix
of an $\MRDlinq{\mu \times (\eta-1), d}$ code, where $\kappa = \eta-d$;
such that the $(\kappa-1)\times(\eta-2)$ submatrix
\begin{equation*}
\Mat{G}_{(\kappa-1) \times (\eta-2)} \defeq
\begin{pmatrix}
1&&& \alpha_{\kappa,1}&\dots & \alpha_{\eta-2,1} \\
&\ddots&& \vdots&\ddots& \vdots\\
&&1& \alpha_{\kappa,\kappa-1}&\dots & \alpha_{\eta-2,\kappa-1}\\
\end{pmatrix}
\end{equation*}
defines an $\MRDlinq{\mu\times (\eta-2), d}$ code;
and such that the $(\kappa-1) \times (\eta-1)$ bottom right submatrix
\begin{equation}\label{eq:genmatk-1n}
\Mat{G}_{(\kappa-1) \times (\eta-1)} \defeq
\left(\begin{array}{c|c}
 &\alpha_{\eta-1,1}\\
\Mat{G}_{(\kappa-1) \times (\eta-2)} &\vdots\\
& \alpha_{\eta-1,\kappa-1}\\
\end{array}\right)
\end{equation}
defines an $\MRDlinq{\mu \times (\eta-1), d+1}$ code.
\end{lemma}

Based on Lemma~\ref{lem:syst_gen_mat}, we can now generalize our construction given in Example~\ref{ex:fourtimesfour_nontriang}.

\begin{construction}[Based on Subcodes of MRD Codes]\label{def:constr_II_subcode}
Let $\delta$ be an integer for which $0 < \delta \leq n-1$.
Let $\ferdigNoInp$ be an $m \times n$ Ferrers diagram with $\gamma_i \geq n-1$, $\forall i \in \intervallexcl{n-\delta+1}{n}$ (i.e., the rightmost $\delta-1$ columns have at least $n-1$ dots). 

Let $\kappa=n-\delta+1$ and let
\begin{equation*}
\Mat{G}_{\kappa \times n} =
\begin{pmatrix}
1&&& &\alpha_{\kappa,0}&\dots & \alpha_{n-2,0} &0\\
&1&&& \alpha_{\kappa,1}&\dots & \alpha_{n-2,1} &\alpha_{n-1,1}\\
&&\ddots&& \vdots&& \vdots\\
&&&1& \alpha_{\kappa,\kappa-1}&\dots & \alpha_{n-2,\kappa-1}& \alpha_{n-1,\kappa-1}\\
\end{pmatrix}
\in \Fqext{n-1}^{\kappa \times n}
\end{equation*}
be a $\kappa \times n$ matrix such that its first $n-1$ columns form a systematic generator matrix of an $\MRDlinq{(n-1) \times (n-1),\delta-1}$ code, and such that the right bottom $(\kappa-1) \times (n-1)$ submatrix (denoted by $\G_{(\kappa-1) \times (n-1)}$) is a systematic generator matrix of an $\MRDlinq{(n-1) \times (n-1),\delta}$ code (constructed as in Lemma~\ref{lem:syst_gen_mat} with $\mu=\eta-1=n-1$ and $d = \delta-1$).

Let the code $\mycodeRank{C}$ be the set of all matrices of the following form:
\begin{equation}
\mycodeRank{C} =
\left\lbrace
\begin{pmatrix}
\\
&&\smash{\clap{\resizebox{1.5cm}{!}{\extsmallfieldinputdeg{\vec{c}}{n-1}}}}\\[4ex]
\hline\\[-2.5ex]
0 &\dots & 0 & u_{0,0}\\
\vdots &\ddots & \vdots & \vdots\\
0 &\dots & 0 & u_{0,s-1}\\
\end{pmatrix}
\in \Fq^{m \times n}
\
:
\
\vec{c} = \vec{u} \cdot \vec{G}_{\kappa \times n} \in \Fqext{n-1}^n, \
\vec{u}\in \Fqext{n-1}^\kappa
\right\rbrace,
\end{equation}
where $\vec{u} = (u_0 \ u_1 \ \dots \ u_{\kappa-1})$ and $\extsmallfieldinputdeg{u_i}{n-1} =  (u_{i,0} \ u_{i,1} \ \dots \ u_{i,n-2})^T$ such that:
\begin{align*}
\extsmallfieldinputdeg{u_0}{n-1} =
\begin{pmatrix}
u_{0,0}\\
\vdots\\
u_{0,\min\{\gamma_0-1,s\}}\\
0\\
\vdots\\
0
\end{pmatrix},
\quad
\extsmallfieldinputdeg{u_i}{n-1} =
\begin{pmatrix}
u_{i,0}\\
\vdots\\
u_{i,\gamma_i-1}\\
0\\
\vdots\\
0
\end{pmatrix},
\quad
u_{0,j},u_{i,j} \in \Fq, \forall i\in \intervallincl{1}{n-\delta}, j\in \intervallexcl{0}{\gamma_i},
\end{align*}
where $s=m-n+1$.
\end{construction}

\begin{lemma}[Properties of Construction~\ref{def:constr_II_subcode}]\label{lem:properties_constrsubcode}
For any $q \geq 2$ and any $m \times n$ Ferrers diagram $\ferdigNoInp$, where the rightmost $\delta-1$ columns have at least $n-1$ dots,
Construction~\ref{def:constr_II_subcode} provides a linear $\ferdigcode{\dimfer}{\delta}$ Ferrers diagram rank-metric code over $\Fq$ of minimum rank distance $\delta$ and
dimension $\dimfer=\min\{s,\gamma_0\}+\sum_{i=1}^{n-\delta} \gamma_i$, where $s=m-n+1$.
\end{lemma}
\begin{IEEEproof}
%
One can easily verify the linearity and the dimension of the code.
It remains to show the minimum rank weight of a non-zero codeword from $\mycodeRank{C}$.
We distinguish between two cases:
\begin{itemize}
\item $u_{0} \neq 0$. The first $n-1$ columns of $\extsmallfieldinputdeg{\vec{c}}{n-1}$, with $\vec{c} = \vec{u} \cdot \Mat{G}_{\kappa\times n}$, constitute
a codeword of an $\MRDlinq{(n-1) \times (n-1), \delta-1}$ code.
Therefore, this $(n-1) \times (n-1)$ submatrix has rank at least $\delta-1$. Hence, the overall rank of such a codeword of $\mycodeRank{C}$ is at least $\delta-1+\rk((u_{0,0} \ u_{0,1} \ \dots \ u_{0,s-1})^T) = \delta$.
\item $u_{0} = 0$. Clearly, $\vec{c} = \big( 0 \quad (u_1 \ u_2 \ \dots \ u_{\kappa-1}) \cdot \Mat{G}_{(\kappa-1) \times (n-1)}\big)$, which is a codeword of an $\MRDlinq{(n-1) \times (n-1), \delta}$ code and hence, the rank of $\extsmallfieldinputdeg{\vec{c}}{n-1}$ is at least $\delta$.
\end{itemize}
\end{IEEEproof}

The following theorem analyzes for which classes of Ferrers diagrams Construction~\ref{def:constr_II_subcode} provides optimal codes.

\begin{theorem}[Optimality of Construction~\ref{def:constr_II_subcode}]\label{thm:optimality_mrd_subcodes}
Let $\ferdigNoInp$ be an $m \times n$ Ferrers diagram and let $\delta$ be an integer, $0 < \delta \leq n-1$, such that
\begin{itemize}
\item the rightmost $\delta-1$ columns of $\ferdigNoInp$ have at least $n-1$ dots,
\item the rightmost column has $\gamma_{n-1}=m= n-1+s \geq n-1+\gamma_0$ dots.
\end{itemize}
Then, for any $q \geq 2$, Construction~\ref{def:constr_II_subcode} provides an optimal Ferrers diagram rank-metric code, i.e., its dimension attains the upper bound from Theorem~1 and its minimum rank distance is $\delta$.
\end{theorem}

\begin{IEEEproof}
Clearly, the upper bound on this type of Ferrers diagrams is obtained by deleting $\delta-1$ columns, and therefore $\dimferdigcode{\delta} \leq \sum_{i=0}^{n-\delta} \gamma_i$. By Lemma~\ref{lem:properties_constrsubcode}, our construction attains this optimal dimension and has minimum rank distance $\delta$.
\end{IEEEproof}

\begin{figure}[htb, scale = 1]
\centering
\includegraphics[scale=1]{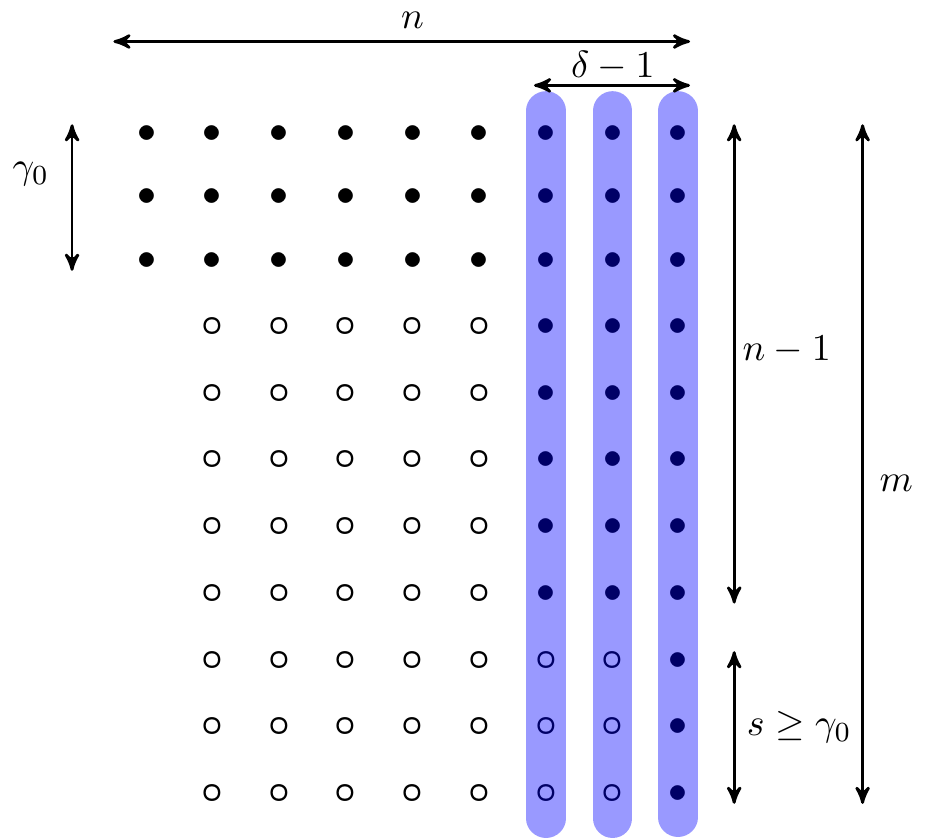}
\caption{Illustration of the Ferrers diagrams, for which Construction~\ref{def:constr_II_subcode} yields optimal codes. The dots "$\bullet$" have to exist, whereas the dots marked by "$\boldsymbol{\circ}$" can exist or not.
Note that the dots in rows two and three of the columns $1, ..., n-\delta$ are "$\bullet$" dots only since the first column has three dots.
\vspace{2ex}
\label{fig:ferrdig_optimalII}}
\hrulefill
\end{figure}

Figure~\ref{fig:ferrdig_optimalII} illustrates Ferrers diagrams for which Construction~\ref{def:constr_II_subcode} provides optimal codes.
Notice that if one of the rightmost $\delta-1$ columns has more than $n-1$ dots, these additional dots are pending dots and therefore the construction does not change. However, if at least one of leftmost $n-\delta+1$ columns has more than $n-1$ dots, then all $\delta-1$ rightmost columns have at least $n$ dots, and an optimal construction is given by \cite{Etzion2009ErrorCorrecting}.

\section{Combining Different Ferrers Diagram Rank-Metric Codes}\label{sec:combination_diagrams}
In this section, we show two possible ways to obtain new Ferrers diagram rank-metric codes based on rank-metric codes in subdiagrams.

First, we combine two Ferrers diagram rank-metric codes of the same dimension.

\begin{theorem}[Combining Codes of the Same Dimension]\label{thm:combine_samedim}
Let $\ferdigNoInp_1$ be an $m_1 \times n_1$ Ferrers diagram and assume $\mycodeRank{C}_1$ is an $\ferdigcodeThreeInp{\ferdigNoInp_1}{\dimfer}{\delta_1}$ code;
let $\ferdigNoInp_2$ be an $m_2 \times n_2$ Ferrers diagram and assume $\mycodeRank{C}_2$ is an $\ferdigcodeThreeInp{\ferdigNoInp_2}{\dimfer}{\delta_2}$ code; let $\mathcal{D}$ be an $m_3 \times n_3$ complete Ferrers diagram (with $m_3 \cdot n_3$ dots), where $m_3 \geq m_1$ and $n_3 \geq n_2$.

Let
\begin{equation*}
\ferdigNoInp =
\left(\begin{matrix}
\ferdigNoInp_1 & \mathcal{D}\\
& \ferdigNoInp_2
\end{matrix}\right),
\end{equation*}
be an $m \times n$ Ferrers diagram $\ferdigNoInp$, where $m =m_2+m_3$ and $n = n_1 +n_3$. Then, there exists an
 $\ferdigcodeThreeInp{\ferdigNoInp}{\dimfer}{\delta_1+\delta_2}$ code.
\end{theorem}
\begin{IEEEproof}
We order $\mycodeRank{C}_1 = \{\Mat{C}^{(1)}_0, \Mat{C}^{(1)}_1,\dots,\Mat{C}^{(1)}_{q^k-1}\}$ and $\mycodeRank{C}_2=\{\Mat{C}^{(2)}_0, \Mat{C}^{(2)}_1,\dots,\Mat{C}^{(2)}_{q^k-1}\}$ such that ${\Mat{C}^{(1)}_i + \Mat{C}^{(1)}_j =\Mat{C}^{(1)}_{\ell}}$ if and only if $\Mat{C}^{(2)}_i + \Mat{C}^{(2)}_j =\Mat{C}^{(2)}_{\ell}$, for all $i,j,\ell$.
Let
\begin{equation*}
\mycodeRank{C} = \left\{\left(\begin{matrix}\Mat{C}^{(1)}_i& \0\\ \0 & \Mat{C}^{(2)}_i\end{matrix}\right):\Mat{C}^{(1)}_i \in \mycodeRank{C}_1, \Mat{C}^{(2)}_i \in \mycodeRank{C}_2 \right\}.
\end{equation*}

Clearly, $\mycodeRank{C}$ is a linear code of dimension $k$. Due to the ordering, $\Mat{C}^{(1)}_i$ and $\Mat{C}^{(2)}_i$ are either both zero or both non-zero. If they are non-zero, then $\rk(\Mat{C}) = \rk(\Mat{C}^{(1)}_i) + \rk(\Mat{C}^{(2)}_i) = \delta_1 + \delta_2$, which proves the minimum rank distance of $\mycodeRank{C}$.
\end{IEEEproof}
The following example shows a diagram in which optimal Ferrers diagram rank-metric codes can be constructed by this strategy. In general, the types of Ferrers diagrams for which we obtain optimal codes with Theorem~\ref{thm:combine_samedim} have some similarity with this example, even if the diagrams can be much larger.
\begin{example}[Combining Codes of the Same Dimension]\label{ex:combin_samedim}
Consider the following diagrams:
\begin{equation*}
\ferdigNoInp = \left(\begin{matrix}
\ferdigNoInp_1 & \mathcal{D}\\
& \ferdigNoInp_2
\end{matrix}\right) = \quad
\begin{matrix}
\bullet & \bullet & \bullet & \bullet \\
\bullet & \bullet & \bullet & \bullet \\
&&&\bullet\\
&&&\bullet\\
&&&\bullet\\
\end{matrix},
\qquad
\ferdigNoInp_1 = \quad
\begin{matrix}
\bullet & \bullet & \bullet\\
\bullet & \bullet & \bullet\\
\end{matrix},
\qquad
\ferdigNoInp_2 = \quad
\begin{matrix}
\bullet\\
\bullet\\
\bullet\\
\end{matrix},
\qquad
\mathcal{D} = \quad
\begin{matrix}
\bullet\\
\bullet\\
\end{matrix}.
\end{equation*}
Assume, we want to construct an optimal rank-metric code of minimum rank distance $\delta=3$ in $\ferdigNoInp$.
A transposed $\MRDlinear{3 \times 2, \delta_1=2}$ code gives an optimal $\ferdigcodeThreeInp{\ferdigNoInp_1}{3}{\delta_1=2}$ code in $\ferdigNoInp_1$.
Further, we consider all vectors in $\Fq^{3 \times 1}$ to obtain an optimal $\ferdigcodeThreeInp{\ferdigNoInp_1}{3}{\delta_2=1}$ code in $\ferdigNoInp_2$.

With Theorem~\ref{thm:combine_samedim}, we can therefore construct an $\ferdigcodeThreeInp{\ferdigNoInp}{3}{\delta_1+\delta_2=3}$ code in $\ferdigNoInp$. This is an optimal code for this diagram since the bound from Theorem~\ref{thm:upper_bound} provides $\dimferdigcode{\delta} \leq 3$.
\end{example}

Next, we combine codes of same minimum rank distance in different Ferrers diagrams.
\begin{theorem}[Combining Codes of Same Distance]\label{thm:combine_samedist}
Let $\ferdigNoInp_1$ be an $m_1 \times n_1$ Ferrers diagram, where the $\ell$ rightmost columns have $m_1$ dots, i.e., $\gamma^{(1)}_{n_1-\ell}= \dots= \gamma^{(1)}_{n_1-1} = m_1$. 
Assume an $\ferdigcodeThreeInp{\ferdigNoInp_1}{\dimfer_1}{\delta}$ code $\mycodeRank{C}_1$ is given.

Further, let $\ferdigNoInp_2$ be an $m_2 \times n_2$ Ferrers diagram, whose first column has $\gamma^{(2)}_0 \geq \gamma^{(1)}_{n_1-\ell-1}$ dots and whose $\ell$ rightmost columns have $m_2$ dots. Assume an $\ferdigcodeThreeInp{\ferdigNoInp_2}{\dimfer_2}{\delta}$ code $\mycodeRank{C}_2$ is given.

\begin{center}
\includegraphics[scale=1]{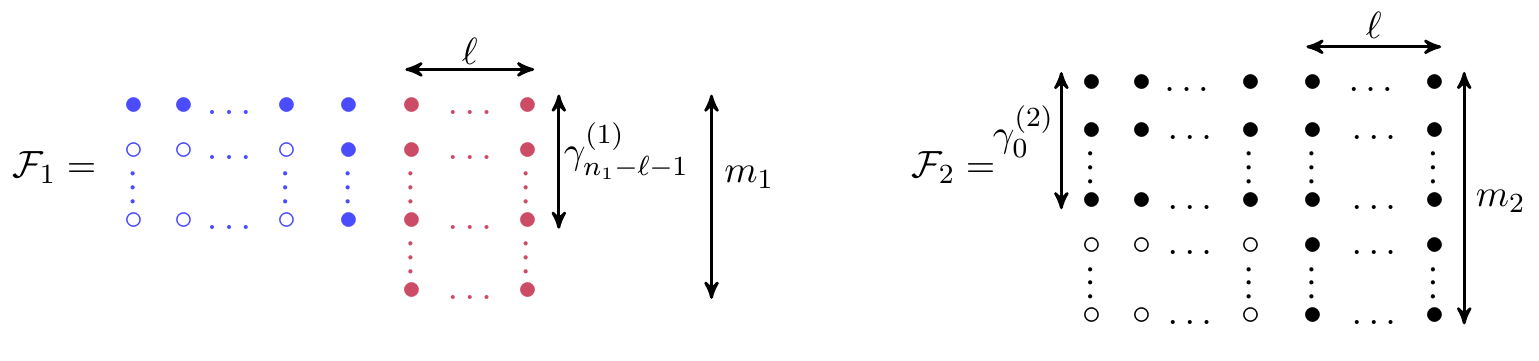}
\end{center}

Then, we can construct an $\ferdigcodeThreeInp{\ferdigNoInp}{\dimfer_1+\dimfer_2}{\delta}$ code $\mycodeRank{C}$ in the following combined $(m_1+m_2) \times (n_1+n_2-\ell)$ Ferrers diagram $\ferdigNoInp$:
\begin{center}
\includegraphics[scale=1]{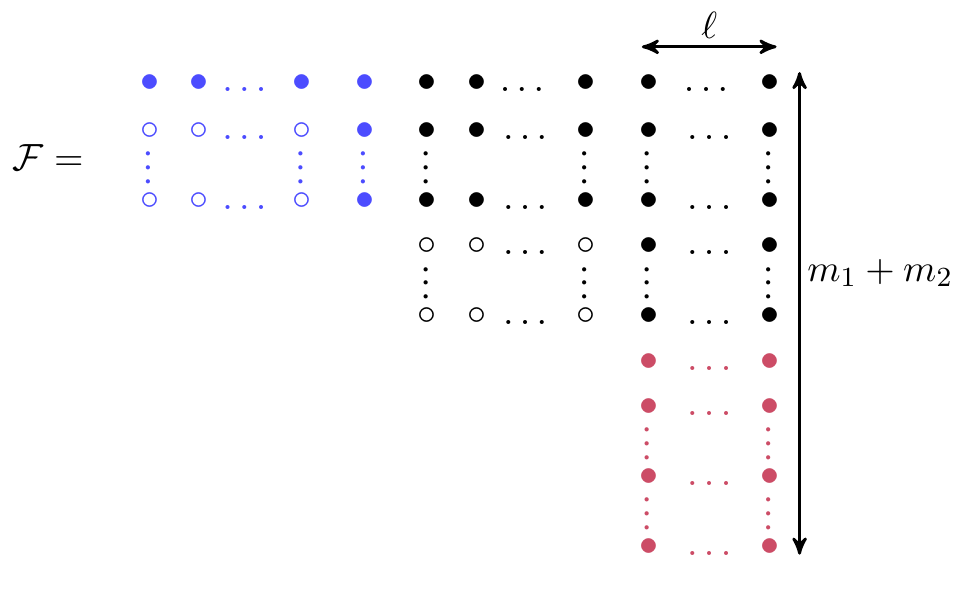}
\end{center}
Further, if both, $\mycodeRank{C}_1$ and $\mycodeRank{C}_2$, attain the upper bound from Theorem~\ref{thm:upper_bound} when $\ell$ columns and $\delta-1-\ell$ rows are deleted in $\ferdigNoInp_1$ and $\ferdigNoInp_2$, then also $\mycodeRank{C}$ attains the upper bound.
\end{theorem}
\begin{IEEEproof}
Let
\begin{equation*}
\mycodeRank{C} = \left\{\begin{pmatrix}
\A & \B\\
\0 & \D
\end{pmatrix}: \B \in \mycodeRank{C}_2, (\A \ \D) \in \mycodeRank{C}_1, \A \in \Fq^{\gamma_{n_1-\ell-1}^{(1)} \times (n_1-\ell)}, \D \in \Fq^{m_1 \times \ell}  \right\}.
\end{equation*}
Clearly, $\mycodeRank{C}$ is a code in $\ferdigNoInp$ of dimension $\dimfer = \dimfer_1+\dimfer_2$. The minimum rank distance of $\mycodeRank{C}$ is $\delta$, since if $\B \neq \0$, the rank of any codeword of $\mycodeRank{C}$ is at least $\delta$ and else, we have to consider $\ferdigNoInp_1$, which is decomposed into two matrices, and since $\rk(\begin{smallmatrix} \Mat{A} & \0 \\ \0 & \Mat{B}\end{smallmatrix}) \geq \rk (\begin{matrix} \Mat{A} & \Mat{B}\end{matrix})$ holds, the minimum rank distance of any non-zero codeword of $\mycodeRank{C}$ is $\delta$.
The upper bound on the dimension is attained for $\ferdigNoInp$ (if it is attained for $\ferdigNoInp_1$ and $\ferdigNoInp_2$) since the same rows and columns as in $\ferdigNoInp_1$ and $\ferdigNoInp_2$ have to be deleted to attain the upper bound on the dimension.
\end{IEEEproof}

\begin{example}[Combining Codes of Same Distance]\label{ex:comb_samedist_nomds}
Consider the following diagrams:
\begin{equation*}
\ferdigNoInp =
\begin{matrix}
\textcolor{blue_aw}{\bullet} &\textcolor{blue_aw}{\bullet} &\textcolor{blue_aw}{\bullet} & \bullet & \bullet & \bullet & \bullet\\
&\textcolor{blue_aw}{\bullet}&\textcolor{blue_aw}{\bullet} & \bullet & \bullet & \bullet & \bullet\\
&&\textcolor{blue_aw}{\bullet} & \bullet & \bullet & \bullet & \bullet\\
&&&&&&\bullet\\
&&&&&&\bullet\\
&&&&&&\bullet\\
&&&&&&\bullet\\
&&&&&&\bullet\\
&&&&&&\bullet\\
&&&&&&\textcolor{institut_color_orig}{\bullet}\\
&&&&&&\textcolor{institut_color_orig}{\bullet}\\
&&&&&&\textcolor{institut_color_orig}{\bullet}\\
&&&&&&\textcolor{institut_color_orig}{\bullet}\\
\end{matrix},
\qquad
\ferdigNoInp_1 = \quad
\begin{matrix}
\textcolor{blue_aw}{\bullet} &\textcolor{blue_aw}{\bullet} & \textcolor{blue_aw}{\bullet} & \textcolor{institut_color_orig}{\bullet}\\
& \textcolor{blue_aw}{\bullet} & \textcolor{blue_aw}{\bullet}&\textcolor{institut_color_orig}{\bullet}\\
&& \textcolor{blue_aw}{\bullet}& \textcolor{institut_color_orig}{\bullet}\\
&&& \textcolor{institut_color_orig}{\bullet}\\
\end{matrix},
\qquad
\ferdigNoInp_2 = \quad
\begin{matrix}
\bullet & \bullet & \bullet & \bullet \\
\bullet & \bullet & \bullet & \bullet \\
\bullet & \bullet & \bullet & \bullet \\
&&&\bullet\\
&&&\bullet\\
&&&\bullet\\
&&&\bullet\\
&&&\bullet\\
&&&\bullet\\
\end{matrix}.
\end{equation*}
Assume, we want to construct an optimal rank-metric code in $\ferdigNoInp$ of minimum rank distance $\delta = 3$.

For all three diagrams, the upper bound on the dimension is attained when deleting one column and one row.
We can construct an optimal  $\ferdigcodeThreeInp{\ferdigNoInp_1}{\dimfer_1=3}{3}$ code in $\ferdigNoInp_1$ with Construction~\ref{def:constr_II_subcode} from Section~\ref{sec:constr_subcode} for any $q \geq 2$.
Further, we can construct an $\ferdigcodeThreeInp{\ferdigNoInp_2}{\dimfer_2 = 6}{3}$ code in $\ferdigNoInp_2$ with the construction from Theorem~\ref{thm:combine_samedim} (similarly as in Example~\ref{ex:combin_samedim}).
This code is optimal since the upper bound on the dimension of any code in $\ferdigNoInp_2$ is also $6$.

The diagrams $\ferdigNoInp_1$ and $\ferdigNoInp_2$ can be combined into $\ferdigNoInp$ (with $\ell=1$) and by Theorem~\ref{thm:combine_samedist}, we can obtain an $\ferdigcodeThreeInp{\ferdigNoInp}{\dimfer_1+\dimfer_2=9}{3}$ code in $\ferdigNoInp$ for any $q \geq 2$.
This is an optimal code for this diagram since the bound from Theorem~\ref{thm:upper_bound} is $\dimferdigcode{\delta} \leq 9$ when deleting one row and one column.

Notice that we can also construct an optimal code for this diagram using Construction~\ref{def:mds_construction} from Section~\ref{sec:construction_mds} with four $\codelinearHamming{4,2,4}$ MDS codes and one $\codelinearHamming{3,1,3}$ MDS code on the diagonals for any $q \geq 3$. However, the construction based on Theorem~\ref{thm:combine_samedist} also provides an optimal code for $q=2$.
\end{example}

\section{Analysis of the Constructions}\label{sec:comparison_andthree}
The previous sections have shown four constructions of Ferrers diagram rank-metric codes (Construction~\ref{def:mds_construction} in Section~\ref{sec:construction_mds}, Construction~\ref{def:constr_II_subcode} in Section~\ref{sec:constr_subcode} and the two possibilities to combine known codes in subdiagrams in Theorems~\ref{thm:combine_samedim} and~\ref{thm:combine_samedist}).
Each of these constructions provides optimal codes for different types of diagrams. In the following, we recall one example for each construction, which cannot be solved by the other constructions and we characterize Ferrers diagrams, for which none of our constructions provides optimal codes.

The following previously shown examples give optimal Ferrers diagram rank-metric codes with the mentioned construction, but none of the others (and also not with \cite{Etzion2009ErrorCorrecting}):
\begin{itemize}
\item Construction~\ref{def:mds_construction}: Example~\ref{ex:constr_mds_one} with $n=m$ and $\delta \geq 4$,
\item Construction~\ref{def:constr_II_subcode}: Example~\ref{ex:fourtimesfour_nontriang} with $\delta=3$,
\item Combination from Theorem~\ref{thm:combine_samedim}: Example~\ref{ex:combin_samedim},
\item Combination from Theorem~\ref{thm:combine_samedist}: Example~\ref{ex:comb_samedist_nomds} with $q=2$.
\end{itemize}
This justifies the existence of each of our four constructions.

Let us now characterize the diagrams for which none of our construction provides optimal codes. Such a diagram has to fulfill all of the following points:
\begin{itemize}
\item it should have at least one element which is not in the first $\ell$ first rows and not in the $\delta - \ell -1$ rightmost columns, and which is below the bottommost diagonal, where $\delta-1$ dots were deleted, of length $s$ (see Fig.~\ref{fig:ferrdig_assumptheorem_mds}) or $q< \numbdots_{max}-1$ such that Construction~\ref{def:mds_construction} does not yield optimal codes,
\item it should have $\gamma_0 > m-n+1$ or at least one of the rightmost $\delta-1$ columns has less than $n-1$ dots such that Construction~\ref{def:constr_II_subcode} does not provide optimal codes,
\item no decompositions as in Theorems~\ref{thm:combine_samedim} and \ref{thm:combine_samedist} should be possible.
\end{itemize}

For $\delta=3$, we believe that we can construct an optimal code for almost all diagrams. We have found no optimal code for $\delta = 3$ and diagrams of the following form:
\begin{center}
\includegraphics[scale=1]{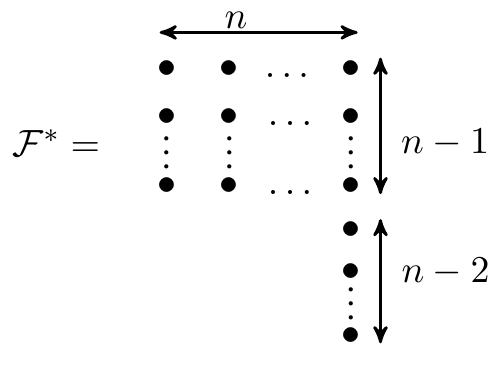}
%
%
%
%
%
\end{center}
However, the following theorem proves that for $\delta=3$, we can construct optimal Ferrers diagram rank-metric codes for any \emph{square} diagram.
\begin{theorem}[Optimal Square Codes for $\delta = 3$]\label{lem:optimal_distance_three}
For any square $n\times n$ Ferrers diagram and any $q \geq 2$, there is an optimal $\ferdigcode{\dimfer}{\delta = 3}$ rank-metric code whose dimension $\dimfer$ attains the upper bound from Theorem~\ref{thm:upper_bound}.
\end{theorem}
\begin{IEEEproof}
As before, denote by $\rho_i$, $\gamma_j$, $i\in \intervallexcl{0}{m}$, $j\in \intervallexcl{0}{n}$, the number of dots in the $i$-th row and $j$-th column, respectively, and distinguish between the following cases:
\begin{itemize}
\item[1)] Assume the upper bound is attained when the \emph{two rightmost columns} are removed (and the bound cannot be attained by removing one row and one column or by removing two rows).
It can easily be verified that in this case $\gamma_{n-2} > \rho_0-1 = n-1$ and the two rightmost columns have exactly $n$ dots. Hence, the construction from Theorem~\ref{thm:es-construction} can be applied and provides optimal codes.

The same clearly holds if two rows are deleted to obtain the upper bound.
\item [2)]Assume the upper bound is attained when \emph{one column and one row} are deleted, i.e.,
\begin{align*}
\gamma_{n-1}+ \rho_0 -1 & \geq  \rho_0 + \rho_1,\\
\gamma_{n-1}+ \rho_0 -1 &\geq\gamma_{n-1}+\gamma_{n-2} .
\end{align*}
Hence, $\rho_1, \gamma_{n-2} \leq n-1$ and therefore $\gamma_0 = 1$.
\begin{itemize}
\item [i)]If $\gamma_{n-2} \geq \rho_1$, consider the top right subdiagram of $\ferdigNoInp$ of size $(\gamma_{n-2}+1) \times (\gamma_{n-2}+1)$, denoted by $\widehat{\ferdigNoInp}$. The Ferrers diagram $\widehat{\ferdigNoInp}$ has $\gamma_{n-2}+1 \leq n$ dots in the first row as well as in the rightmost column; its upper bound on the dimension is the same as the one for $\ferdigNoInp$, since no dots which were not in the first row or rightmost column of $\ferdigNoInp$ were deleted and the number of dots in $\widehat{\ferdigNoInp}$ in the first row and the rightmost column is $2 \gamma_{n-2}+1$, the number of dots in the two rightmost columns is also $2\gamma_{n-2} +1$ and in the two top rows the number of dots is $\gamma_{n-2}+1 + \rho_1 \leq 2 \gamma_{n-2}+1 $. Therefore, in $\widehat{\ferdigNoInp}$, we also obtain the upper bound when deleting two columns. Since $\gamma_0=1$, we can apply Construction~\ref{def:constr_II_subcode} from Definition~\ref{def:constr_II_subcode} with $s=1$ on this subdiagram and obtain optimal codes.
\item[ii)] Else if $\gamma_{n-2} < \rho_1$, in the same way, we can apply Construction~\ref{def:constr_II_subcode} on the transpose of the top right subdiagram of size $(\rho_1+1) \times (\rho_1+1)$.
%
\end{itemize}
\end{itemize}
Therefore, in any case, we can construct an optimal $\ferdigcode{\dimfer}{\delta = 3}$ code in any square Ferrers diagram.
\end{IEEEproof}

\section{Conclusion and Outlook}
\label{sec:conclusion}

We have presented four constructions of rank-metric codes in Ferrers diagrams and we have proven for which diagrams these constructions provide optimal such codes.
Each of our four possible constructions matches a different type of diagrams, i.e., they give optimal codes for different patterns of dots. One construction is based on MDS codes, one on subcodes of MRD codes and two are combinations of smaller codes.

For future work, we suggest the following open questions:
\begin{itemize}
\item Construction~\ref{def:mds_construction} works only when the field size $q$ is sufficiently large.
Can we give another construction which provides optimal codes for the same types of diagrams, but for any $q \geq 2$?
\item Find optimal code constructions for $\delta=3$ for diagrams like $\ferdigNoInp^*$ at the end of Section~\ref{sec:comparison_andthree}, e.g., for $\delta=3$ and the following $5 \times 4$ Ferrers diagram:
\begin{equation*}
\begin{matrix}
\bullet &\bullet &\bullet &\bullet \\
\bullet &\bullet &\bullet &\bullet \\
\bullet &\bullet &\bullet &\bullet \\
&&&\bullet\\
&&&\bullet\\
\end{matrix}.
\end{equation*}
This implies the question: can the bound from Theorem~\ref{thm:upper_bound} be attained for all Ferrers diagrams when $\delta=3$?
\item Find optimal code constructions for arbitrary $\delta$ and diagrams, which are not covered by any of our constructions. Such an example is the following diagram with $\delta=4$:
\begin{equation*}
\begin{matrix}
\bullet & \bullet & \bullet & \bullet& \bullet & \bullet\\
\bullet & \bullet & \bullet & \bullet& \bullet & \bullet\\
&& \bullet & \bullet & \bullet & \bullet\\
&& \bullet &\bullet& \bullet & \bullet\\
&&&& \bullet& \bullet\\
&&&& \bullet& \bullet\\
\end{matrix}.
\end{equation*}
Therefore: are there parameters for which the bound from Theorem~\ref{thm:upper_bound} cannot be attained or is the bound always tight?


\item Can we use cyclically continued MDS codes in diagrams? Consider the diagram from Example~\ref{ex:fourtimesfour_nontriang} with $\delta = 3$:
\begin{equation*}
\begin{matrix}
\bullet &\bullet &\bullet &\bullet \\
 &\bullet &\bullet &\bullet \\
 &\bullet &\bullet &\bullet \\
&&&\bullet\\
\end{matrix}.
\end{equation*}
Can we construct an optimal (i.e., $k = 4$) Ferrers diagram rank-metric code by using an $\codelinearHamming{4,2,3}$ MDS code on $\diagonal_3$, an $\codelinearHamming{3,1,3}$ MDS code on $\diagonal_2$, and \emph{additionally} an
$\codelinearHamming{3,1,3}$ MDS code on the three points of $\diagonal_1$ and~$\diagonal_4$? The difficulty of such a construction is to prove the minimum rank distance.

\item One interesting case are $n \times n$ Ferrers diagram rank-metric codes.
For these diagrams the bound of Theorem~\ref{thm:upper_bound} is attained for
$\delta=2$~\cite{Etzion2009ErrorCorrecting} and for $\delta=3$
(see Theorem~\ref{lem:optimal_distance_three}). Some cases for $\delta=n$ were
considered in~\cite{GorlaRavagnani-SubspaceCodesFerrersDiagrams_2014}
(one of these case can be solved by Theorem~\ref{thm:es-construction}; the
second case and other diagrams for $\delta =n$ can be solved by
Theorem~\ref{thm:combine_samedim}).
We would like to see this case solved for all distances.

\item Finally, one can ask how close can we get to the upper bound
of Theorem~\ref{thm:upper_bound}. It is not difficult to prove that
we can obtain a code of dimension within $(\delta -1)n$ of the upper
bound with the known constructions, but we think that this is a weak result
and believe that it can be significantly
improved with the known constructions.

\end{itemize}

\vspace{1ex}
\begin{center}
{\bf Note added}
\end{center}
After our results were completed (see~arxiv:1405.1885v1) another
submission (see~arxiv:1405.2736v1~\cite{GorlaRavagnani-SubspaceCodesFerrersDiagrams_2014}), which was done independently
and which is focused on a different angle for
similar problems, appeared. Both submissions share
the construction based on the MDS codes (Construction~\ref{def:mds_construction}).

\section*{Appendix}
\begin{IEEEproof}[Proof of Lemma~\ref{lem:syst_gen_mat}]
Let $g_0 = 1,g_1,\dots,g_{\eta-2} \in \Fqext{\mu}$ be linearly independent over $\Fq$. Then, for $\kappa = \eta-d$, the following matrix defines a $\Gab{\mu \times (\eta-1), d}$ code in vector representation over $\Fqext{\mu}$ (see \cite{Gabidulin_TheoryOfCodes_1985}):
\begin{equation*}
\Mat{G}_{0} =
\begin{pmatrix}
1 & g_1 & \dots & g_{\eta-2}\\
1 & g_1^{[1]} & \dots & g_{\eta-2}^{[1]}\\
\vdots& \vdots & \ddots & \vdots\\
1 & g_1^{[\kappa-1]} & \dots & g_{\eta-2}^{[\kappa-1]}\\
\end{pmatrix}.
\end{equation*}
For any full-rank matrix $\Mat{T} \in \Fqext{\mu}^{\kappa \times \kappa}$, the generator matrix $\Mat{T} \cdot \Mat{G}_{0}$ defines the same code as $\Mat{G}_{0}$. Hence, let $t_{0,1}, \dots, t_{0,\kappa-1} \in \Fqext{\mu}$ be such that
\begin{align*}
\Mat{G}_{1}
&=\begin{pmatrix}
1 & t_{0,1} & \dots & t_{0,\kappa-1}\\
& 1\\
& &\ddots\\
& &&1\\
\end{pmatrix}
\begin{pmatrix}
1\\
-1 & 1\\
\vdots && \ddots\\
-1 && &1\\
\end{pmatrix}
\cdot \Mat{G}_{0}
\\
&=
\begin{pmatrix}
1 & 0 & \dots & 0&\alpha_{\kappa,0}&\dots & \alpha_{\eta-2,0}\\
0 & g_1^{[1]} - g_1 & \dots & g_{\kappa-1}^{[1]} - g_{\kappa-1} &  g_{\kappa}^{[1]} - g_{\kappa} & \dots&   g_{\eta-2}^{[1]} - g_{\eta-2}\\
0 & g_1^{[2]} - g_1 &\dots & g_{\kappa-1}^{[2]} - g_{\kappa-1} &  g_{\kappa}^{[2]} - g_{\kappa} &\dots&   g_{\eta-2}^{[2]} - g_{\eta-2}\\
\vdots& \vdots& \ddots &\vdots  & \vdots &\ddots & \vdots\\
0 & g_1^{[\kappa-1]} - g_1 & \dots & g_{\kappa-1}^{[\kappa-1]} - g_{\kappa-1} & g_{\kappa}^{[\kappa-1]} - g_{\kappa} & \dots&  g_{\eta-2}^{[\kappa-1]} - g_{\eta-2}\\
\end{pmatrix}.
\end{align*}
Notice that $t_{0,1}, \dots, t_{0,\kappa-1}$ influence only the first row of $\Mat{G}_{1}$ and the requirements on the first row constitute a heterogeneous linear system of equations with $\kappa-1$ equations and $\kappa-1$ unknowns.
Therefore, such entries $t_{0,1}, \dots, t_{0,\kappa-1}$ always exist.
Further, let
\begin{align*}
\Mat{G}_{2} &=
\begin{pmatrix}
1 \\
& 1\\
& -1& 1\\
 &&\ddots&\ddots\\
&&&-1&1\\
\end{pmatrix}
\Mat{G}_{1}\\
&=
\begin{pmatrix}
1 & 0 &  \dotsm & 0 & \alpha_{\kappa,0} &\dots &\alpha_{\eta-2,0}\\
0 & g_1^{[1]} - g_1  & \dots & g_{\kappa-1}^{[1]} - g_{\kappa-1}& g_{\kappa}^{[1]} - g_{\kappa}&\dots & g_{\eta-2}^{[1]} - g_{\eta-2}\\
0 & (g_1^{[1]} - g_1)^{[1]} &  \dots& (g_{\kappa-1}^{[1]} - g_{\kappa-1})^{[1]}& (g_{\kappa}^{[1]} - g_{\kappa})^{[1]} &\dots &(g_{\eta-2}^{[1]} - g_{\eta-2})^{[1]}\\
\vdots& \vdots&  \ddots & \vdots& \vdots&  \ddots & \vdots\\
0 & (g_1^{[1]} - g_1)^{[\kappa-1]}  &\dots &(g_{\kappa-1}^{[1]} - g_{\kappa-1})^{[\kappa-1]}& (g_{\kappa}^{[1]} - g_{\kappa})^{[\kappa-1]}& \dots & (g_{\eta-2}^{[1]} - g_{\eta-2})^{[\kappa-1]}\\
\end{pmatrix}.
\end{align*}
Since we have only multiplied several times by full-rank matrices from the left, $\Mat{G}_{2}$ defines the same $\Gab{\mu \times (\eta-1), d}$ code as $\Mat{G}_{0}$.
Since $g_1,g_2, \dots, g_{\eta-2}$ are linearly independent over $\Fq$, also $g_1^{[1]}-g_1,g_2^{[1]}-g_2, \dots, \ g_{\eta-2}^{[1]}-g_{\eta-2} \in \Fqext{\mu}$ are linearly independent\footnote{In fact this is a codeword of the $\MRDlinq{\mu \times (\eta-1), \eta-2}$ code, which is generated by $\left(\begin{smallmatrix} 1 & g_1 & \dots & g_{\eta-2}\\ 1 & g_1^{[1]} & \dots & g_{\eta-2}^{[1]}\end{smallmatrix}\right)$ and thus $\rk\left(\extsmallfieldinputdeg{g_1^{[1]}-g_1 \ \  g_2^{[1]}-g_2\ \dots \ g_{\eta-2}^{[1]}-g_{\eta-2}}{\mu}\right) = \eta-1$.} over $\Fq$ and the right bottom $(\kappa-1) \times (\eta-2)$ submatrix of $\Mat{G}_{2}$ defines an $\Gab{\mu \times (\eta-2), d}$ code. Additionally, since $\mu \geq \eta-1$, there exists an element $g_{\eta-1} \in \Fqext{\mu}$ which is $\Fq$-linearly independent of $g_1^{[1]}-g_1, \dots, g_{\eta-2}^{[1]}-g_{\eta-2} $.
Hence, the $\kappa \times \eta$ matrix
\begin{equation*}
\left(\begin{array}{c|c}
& 0\\
& g_{\eta-1}\\
\Mat{G}_{2}& g_{\eta-1}^{[1]}\\
& \vdots\\
& g_{\eta-1}^{[k-1]}
\end{array}\right),
\end{equation*}
defines with its first $\eta-1$ columns (due to ${\Mat{G}}_{2}$) a $\Gab{\mu \times (\eta-1), d}$ code and with the right bottom $(\kappa-1) \times (\eta-1)$ submatrix a $\Gab{\mu \times (\eta-1), d+1}$ code.

Finally, we can choose an invertible matrix $\Mat{T}_{\kappa-1} \in \Fqext{\mu}^{(\kappa-1) \times (\kappa-1)}$ such that we obtain the systematic generator matrix $\G_{\kappa \times \eta}$ from the statement:
\begin{equation*}
\G_{\kappa \times \eta}
=
\begin{pmatrix}
1 \\
& \Mat{T}_{\kappa-1}\\
\end{pmatrix}
\cdot
\left(\begin{array}{c|c}
& 0\\
& g_{\eta-1}\\
\Mat{G}_{2}& g_{\eta-1}^{[1]}\\
& \vdots\\
& g_{\eta-1}^{[\kappa-1]}
\end{array}\right).
\end{equation*}
Since $\Mat{T}_{\kappa-1}$ performs linear combinations only of the $\kappa-1$ lower rows, $\Mat{G}_{\kappa \times \eta}$ is a systematic matrix with the properties of the statement.
\end{IEEEproof}

\bibliographystyle{IEEEtranS}
\bibliography{antoniawachter}

\begin{thebibliography}{10}
\providecommand{\url}[1]{#1}
\csname url@samestyle\endcsname
\providecommand{\newblock}{\relax}
\providecommand{\bibinfo}[2]{#2}
\providecommand{\BIBentrySTDinterwordspacing}{\spaceskip=0pt\relax}
\providecommand{\BIBentryALTinterwordstretchfactor}{4}
\providecommand{\BIBentryALTinterwordspacing}{\spaceskip=\fontdimen2\font plus
\BIBentryALTinterwordstretchfactor\fontdimen3\font minus
  \fontdimen4\font\relax}
\providecommand{\BIBforeignlanguage}[2]{{%
\expandafter\ifx\csname l@#1\endcsname\relax
\typeout{** WARNING: IEEEtranS.bst: No hyphenation pattern has been}%
\typeout{** loaded for the language `#1'. Using the pattern for}%
\typeout{** the default language instead.}%
\else
\language=\csname l@#1\endcsname
\fi
#2}}
\providecommand{\BIBdecl}{\relax}
\BIBdecl

\bibitem{AndrewsEriksson-IntegerPartitions_Book}
G.~E. Andrews and K.~Eriksson, \emph{{Integer partitions}}, 2nd~ed.\hskip 1em
  plus 0.5em minus 0.4em\relax Cambridge University Press, Oct. 2004.

\bibitem{Bachoc2012Bounds}
C.~Bachoc, F.~Vallentin, and A.~Passuello, ``{Bounds for projective codes from
  semidefinite programming},'' \emph{Adv. Math. Commun.}, vol.~7, no.~2, pp.
  127--145, May 2013.

\bibitem{Delsarte_1978}
P.~Delsarte, ``{Bilinear forms over a finite field with applications to coding
  theory},'' \emph{J. Combin. Theory Ser. A}, vol.~25, no.~3, pp. 226--241,
  1978.

\bibitem{Etzion2009ErrorCorrecting}
T.~Etzion and N.~Silberstein, ``{Error-correcting codes in projective spaces
  via rank-metric codes and Ferrers diagrams},'' \emph{IEEE Trans. Inform.
  Theory}, vol.~55, no.~7, pp. 2909--2919, Jul. 2009.

\bibitem{EtzionSilberstein-CodesDesignsRelLiftedMRD-2012}
------, ``{Codes and designs related to lifted MRD codes},'' \emph{IEEE Trans.
  Inform. Theory}, vol.~59, pp. 1004--1017, Feb. 2013.

\bibitem{Etzion2011ErrorCorrecting}
T.~Etzion and A.~Vardy, ``{Error-correcting codes in projective space},''
  \emph{IEEE Trans. Inform. Theory}, vol.~57, no.~2, pp. 1165--1173, Feb. 2011.

\bibitem{Gabidulin_TheoryOfCodes_1985}
E.~M. Gabidulin, ``{Theory of codes with maximum rank distance},'' \emph{Probl.
  Inf. Transm.}, vol.~21, no.~1, pp. 3--16, 1985.

\bibitem{GabidulinPilipchuk-RankSubcodesMulticomponentNetworkCoding_2013}
E.~M. Gabidulin and N.~I. Pilipchuk, ``{Rank subcodes in multicomponent network
  coding},'' vol.~49, no.~1, pp. 40--53, 2013.

\bibitem{Gadouleau2010ConstantRank}
M.~Gadouleau and Z.~Yan, ``{Constant-rank codes and their connection to
  constant-dimension codes},'' \emph{IEEE Trans. Inform. Theory}, vol.~56,
  no.~7, pp. 3207--3216, Jul. 2010.

\bibitem{GorlaRavagnani-SubspaceCodesFerrersDiagrams_2014}
\BIBentryALTinterwordspacing
E.~Gorla and A.~Ravagnani, ``{Subspace codes from Ferrers diagrams},'' May
  2014. [Online]. Available: \url{http://arxiv.org/abs/1405.2736}
\BIBentrySTDinterwordspacing

\bibitem{KhKs09}
\BIBentryALTinterwordspacing
A.~Khaleghi and F.~R. Kschischang, ``{Projective space codes for the injection
  metric},'' Feb. 2009. [Online]. Available:
  \url{http://arxiv.org/abs/0904.0813}
\BIBentrySTDinterwordspacing

\bibitem{KhaleghiSilvaKschischang-SubspaceCodes_2009}
A.~Khaleghi, D.~Silva, and F.~R. Kschischang, ``{Subspace codes},'' in
  \emph{Cryptography and Coding}, ser. Lecture Notes in Computer Science, 2009,
  vol. 5921, pp. 1--21.

\bibitem{KohnertKurz-LargeConstantDimensionCodes-2008}
A.~Kohnert and S.~Kurz, ``{Construction of large constant dimension codes with
  a prescribed minimum distance},'' in \emph{Mathematical Methods in Computer
  Science}, ser. Lecture Notes in Computer Science.\hskip 1em plus 0.5em minus
  0.4em\relax Springer Berlin Heidelberg, 2008, vol. 5393, pp. 31--42.

\bibitem{koetter_kschischang}
R.~K\"otter and F.~R. Kschischang, ``{Coding for errors and erasures in random
  network coding},'' \emph{IEEE Trans. Inform. Theory}, vol.~54, no.~8, pp.
  3579--3591, Jul. 2008.

\bibitem{MacWilliamsSloane_TheTheoryOfErrorCorrecting_1988}
F.~J. MacWilliams and N.~J.~A. Sloane, \emph{{The theory of error-correcting
  codes}}.\hskip 1em plus 0.5em minus 0.4em\relax North Holland Publishing Co.,
  1988.

\bibitem{Roth_RankCodes_1991}
R.~M. Roth, ``{Maximum-rank array codes and their application to crisscross
  error correction},'' \emph{IEEE Trans. Inform. Theory}, vol.~37, no.~2, pp.
  328--336, 1991.

\bibitem{SilbersteinTrautmann-NewLowerBoundsForConstantDimensionCodes_2013}
N.~Silberstein and A.~L. Trautmann, ``{New lower bounds for constant dimension
  codes},'' in \emph{IEEE Int. Symp. on Inf. Theory}, Jul. 2013, pp. 514--518.

\bibitem{SilbersteinTrautmann-SubspaceCodesBasedOnGraphmatchingFerrersDig_2014}
\BIBentryALTinterwordspacing
N.~Silberstein and A.-L. Trautmann, ``{Subspace codes based on graph matchings,
  Ferrers diagrams and pending blocks},'' Apr. 2014. [Online]. Available:
  \url{http://arxiv.org/abs/1404.6723}
\BIBentrySTDinterwordspacing

\bibitem{SilvaKschischang-MetricsErrorCorrectionNetworkCoding_2009}
D.~Silva and F.~R. Kschischang, ``{On metrics for error correction in network
  coding},'' \emph{IEEE Trans. Inform. Theory}, vol.~55, no.~12, pp.
  5479--5490, Dec. 2009.

\bibitem{silva_rank_metric_approach}
D.~Silva, F.~R. Kschischang, and R.~K{\"o}tter, ``{A rank-metric approach to
  error control in random network coding},'' \emph{IEEE Trans. Inform. Theory},
  vol.~54, no.~9, pp. 3951--3967, 2008.

\bibitem{Skachek2010Recursive}
V.~Skachek, ``{Recursive code construction for random networks},'' \emph{IEEE
  Trans. Inform. Theory}, vol.~56, no.~3, pp. 1378--1382, Mar. 2010.

\bibitem{TrautmannManganielloRosenthal-OrbitCodes-2010}
A.~L. Trautmann, F.~Manganiello, and J.~Rosenthal, ``{Orbit codes - a new
  concept in the area of network coding},'' in \emph{IEEE Information Theory
  Workshop 2019 (ITW 2012)}, Aug. 2010.

\bibitem{TrautmannRosenthal-ImprovementsEchelonFerrersConstruction_2010}
A.~L. Trautmann and J.~Rosenthal, ``{New improvements on the Echelon-Ferrers
  construction},'' in \emph{19th International Symposium on Mathematical Theory
  of Networks and Systems (MTNS)}, Jul. 2010, pp. 405--408.

\bibitem{Wang2003Linear}
H.~Wang, C.~Xing, and R.~{Safavi-Naini}, ``{Linear authentication codes: bounds
  and constructions},'' \emph{IEEE Trans. Inform. Theory}, vol.~49, no.~4, pp.
  866--872, Apr. 2003.

\bibitem{Xia2009Johnson}
S.~Xia and F.~Fu, ``{Johnson type bounds on constant dimension codes},''
  \emph{Des. Codes Cryptogr.}, vol.~50, no.~2, pp. 163--172, Feb. 2009.

\end{thebibliography}
\end{document}